\documentclass[12pt]{iopart}
\usepackage{iopams}  
\usepackage{setstack} 
\usepackage{cite}
\usepackage{url}
\usepackage{graphicx}
\usepackage{leftidx}
\usepackage{xcolor}
\usepackage{soul}
\usepackage{bm}
\usepackage{subfigure}
\newcommand{\Pe}{{\rm Pe}}

\usepackage[colorlinks=true, allcolors=blue]{hyperref}

\newcommand{\calQ}{{\cal Q}}

\newcommand{\calX}{{\cal X}}
\newcommand{\calB}{{\cal B}}
\newcommand{\calM}{{\cal M}}
\newcommand{\calY}{{\cal Y}}
\newcommand{\calZ}{{\cal Z}}

\newcommand{\hu}{\hat{\bm u}}
\newcommand{\kapt}{\tilde{\kappa}}
\newcommand{\mycomment}[1]{}

\begin{document}

\title{Dissipation-accuracy tradeoffs in autonomous control of smart active matter}

\author{Luca Cocconi\textsuperscript{1}, Beno{\^i}t Mahault\textsuperscript{1} and Lorenzo Piro\textsuperscript{1,2}}
\address{$^1$ Max Planck Institute for Dynamics and Self-Organization (MPIDS), 37077 G{\"o}ttingen, Germany\\
$^2$ Department of Physics \& INFN, University of Rome ``Tor Vergata", Via della Ricerca Scientifica 1, 00133 Rome, Italy\\}

\ead{luca.cocconi@ds.mpg.de}
\vspace{10pt}
\begin{indented}
\item[]\today
\end{indented}

\begin{abstract}
The study of motility control by smart agents offers a promising platform for systematically exploring the fundamental physical constraints underlying the functioning of bio-inspired micro-machines operating far from equilibrium. Here, we address the question of the energy cost required for a self-steering active agent to localise itself within a specific region of space or follow a pre-defined trajectory under the influence of fluctuations and external flows. Building on a stochastic thermodynamic formulation of the problem, we derive a generic relationship between dissipation and localisation accuracy, which reveals a fundamental dissipation-accuracy tradeoff constraining the agent's performance. In addition, we illustrate how our framework enables the derivation of optimal steering policies that achieve localisation at minimum energy expenditure.
\end{abstract}

\section{Introduction}

Adaptability to a changing environment is often set as a criterion to define “alive-ness”. 
Regulatory feedbacks in response to environmental cues are indeed a central concept in biology as they enable homeostasis, while they are also involved in a variety of processes such as chemotaxis, circadian rhythms, or morphogenesis \cite{milo2002network,alon2019introduction}. 
At the same time, while not bound by equilibrium thermodynamics, living matter is still energetically constrained by the resources available to it \cite{yang2021physical,the_economic_cell_collective_2023_8156387}, leading to quality-dissipation bounds on sensing \cite{bo2015thermodynamic,lan2012energy,govern2014optimal,OwenPRX2020}, signal transmission \cite{Bryant2023}, error correction \cite{sartori2015thermodynamics}, or biological discrimination \cite{yu2022energy}. 
Characterizing the energetics of regulatory feedbacks is therefore of obvious interest to further our understanding of the fundamental capabilities and trade-offs underlying the functioning of biological matter, while it may also find application to the design of biomimetic materials with robust functions.

Stochastic thermodynamics has provided a number of theoretical insights on this question.
For example, the discovery of the thermodynamic uncertainty relation (TUR) \cite{barato2015thermodynamic,horowitz2020thermodynamic,dieball2023direct}, which relates the fluctuations of time-extensive currents to the overall dissipation in a non-equilibrium steady state, has contributed to energizing a wave of still on-going research into the fundamental physical limits imposed on the operation of living and synthetic machines at the microscale \cite{brown2019theory,PhysRevE.101.062123,pietzonka2016universal,ouldridge2017thermodynamics,hwang2018energetic}. 
In a number of relevant cases, however, regulation is not obviously associated with time-extensive currents and thus falls outside the scope of the TUR.

One such case has to do with the navigation of microorganisms capable of processing information about their environment to optimize their search for nutrients, mating partners, or escape predators~\cite{LaugaRev2016,Tso1974}.
Common strategies employed by bacteria, for instance, rely on tactic behavior making use of local chemical or light gradients.
In a complex environment, efficient navigation of microswimmers may also be achieved when they seek to follow pre-determined trajectories that take into account the influence of the flow landscape~\cite{Liebchen_2019,DaddiComPhys2021,piro2022efficiency,piro2022optimal,piro2024energetic}.
In this context, regulatory feedbacks can be employed to overcome Brownian motion and confine the trajectory of the swimmer to a target subspace~\cite{piro2022efficiency,piro2023optimal}, which necessarily comes at a cost and thus raises the multi-objective optimization problem of balancing precision and energy expenditure.
While much is known about the energetics of motility generation~\cite{brown2019theory,DaddiNatCom2023} and notwithstanding recent calls to integrate concepts of control theory into the study of active matter \cite{levine2023physics,goldman2024robot}, the thermodynamic constraints on the information processing required by smart active agents to navigate efficiently remain comparatively less explored~\cite{nava2018markovian,paoluzzi2020information,khadka2018active,muinos2021reinforcement,stark2021artificial,schneider2019optimal}.

\begin{figure}
    \centering
    \includegraphics[scale=0.7]{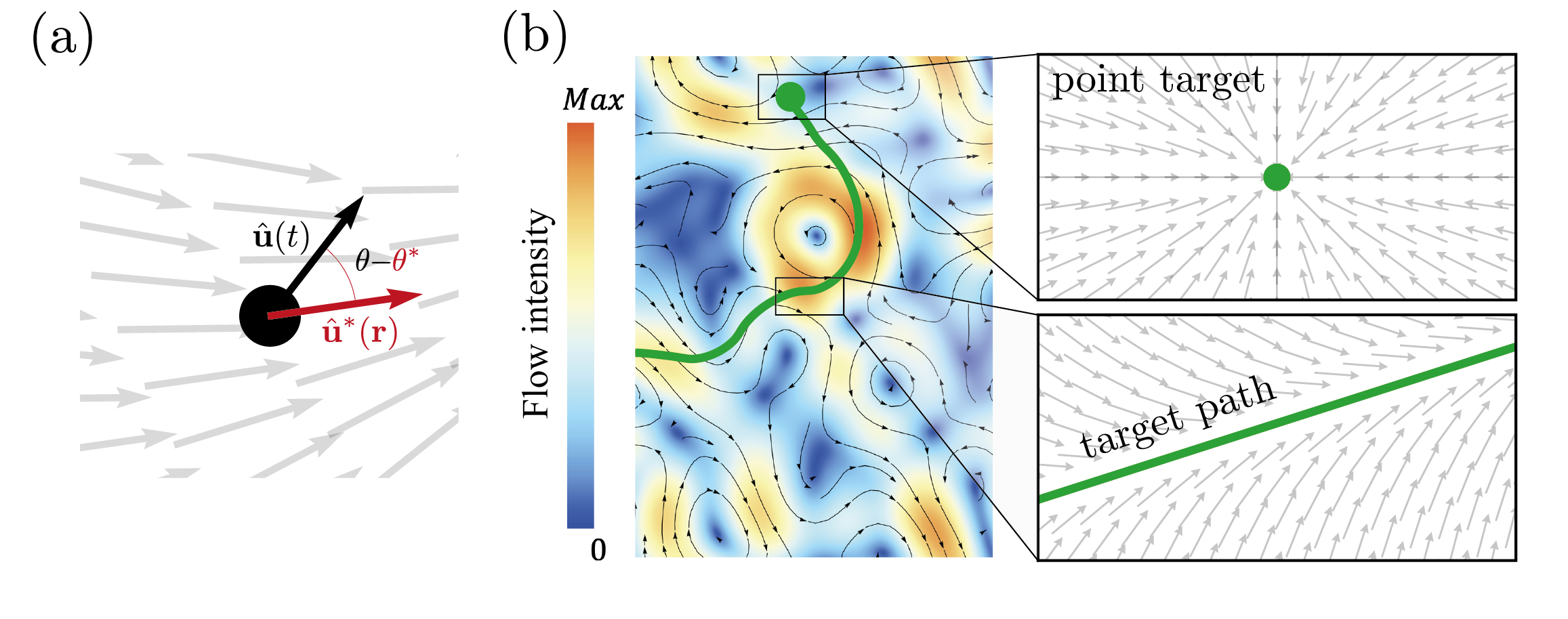}
    \caption{
    Navigation by a self-steering active Brownian particle.
    (a) The agent is subject to a self-generated torque attempting to align its instantaneous self-propulsion direction (black vector) with a given position-dependent steering policy (grey vector field). In addition, the agent is subject to both positional and angular diffusion with thermal origin, as described in Eqs.~(\ref{eq:lang_r},\ref{eq:lang_t}). 
    (b) In this work, we consider steering policies designed to localise self-steering agents to a lower dimensional target subspace (green). For example, this subspace may correspond to a point target or to a desired trajectory between two points in a complex flow field (target path). 
    Grey vectors in the zoom-ins show examples of possible policies.
    }
    \label{fig:intro_schematic}
\end{figure}

Here, we establish a thermodynamically consistent framework for the study of motility control by an agent moving at constant speed and capable of autonomous steering. Using navigation as a case study, we explore steering policies that correct for external positional and rotational noises, with the objective of actively confining the swimmer to a specific region or trajectory in space, such as a point target or a line (see Fig.~\ref{fig:intro_schematic} for an illustration). An immediate benefit of treating motility and control in a unified framework is the possibility of decomposing the ensuing dissipation —quantified via the mean rate of entropy production~\cite{seifert2012stochastic,cocconi2020entropy}— into distinct contributions related to self-propulsion and steering, respectively. Our analysis reveals a fundamental dissipation-accuracy tradeoff, which we use to quantify the increase in dissipation associated with steering required to improve the localisation of the swimmer. We highlight the generality of this result in various scenarios, including localisation at a point target or along a target line, with and without the presence of an external flow advecting the swimmer. In the absence of flow, we further identify a family of optimal policies for which better precision can only be achieved at the cost of additional dissipation, thus defining the Pareto front~\cite{ngatchou2005pareto} associated with this multi-objective optimization problem.\\

The paper is organized as follows: we start by introducing a general self-steering swimmer model in Section~\ref{s:model_setup}, before focusing on the weak alignment regime where we derive an effective dynamics for the swimmer density, which enables an analytical exploration of the problem.
Drawing on these results, we consider a minimal geometry in Section~\ref{s:pt_target}, where the steering policy aims to localise the agent at a point target, compensating for the effects of diffusion and external flows. This analysis demonstrates the existence of a generic dissipation-accuracy tradeoff, which we extend in Section~\ref{s:target_path} to policies aimed at localising the agent in the proximity of a one-dimensional target path.
In the simplest case, where external flows are absent, we determine a family of optimal policies that overlap with the Pareto front. 
We then discuss how these results are affected by the presence of non-trivial flows.
We summarise our findings in Section~\ref{s:discussion_conclusion} and provide a perspective on future research directions. Additional results and details on the analytical derivations are presented in the appendices.

\section{Model setup} \label{s:model_setup}

\subsection{Active Brownian motion with feedback control}

We consider a self-propelled agent moving in two dimensions at constant speed $w$ in the presence of a stationary flow field $\bm{v}(\bm{r})$.
The swimmer is characterised by its center of mass position $\bm{r}(t)$ and heading direction $\hat{\bm{u}}(\theta) = (\cos\theta,\sin\theta)$, with $\theta \in [0,2\pi)$. 
Circumflexes will henceforth indicate unit vectors. The dynamics of the position and heading angle are governed by the following 
pair of overdamped Langevin equations
\begin{eqnarray}
    \dot{\bm{r}}(t) &= w \hat{\bm{u}}(\theta) + \bm{v}(\bm{r}) + \sqrt{2D_t} \bm{\xi}(t) , \label{eq:lang_r}\\
    \dot{\theta}(t) &= \kappa \Gamma(\theta,\bm{r})+ \sqrt{2D_\theta} \xi_\theta(t) , \label{eq:lang_t}
\end{eqnarray}
where 
the $\xi_{i}$ ($i\in \{x,y,\theta\}$) are independent and identically distributed Gaussian white noises of unit covariance, while $D_t$ and $D_\theta$ denote the translational and angular diffusion coefficients, respectively. 
The angular velocity $\kappa \Gamma$ originates from a self-generated torque that allows the agent to actively control its moving direction.
For $\kappa = 0$, Eqs.~(\ref{eq:lang_r},\ref{eq:lang_t}) reduce the description of an active Brownian particle (ABP) in two dimensions, a paradigmatic model of active matter~\cite{bechinger2016active}.
Here, we take
\begin{equation}
    \Gamma(\theta,\bm{r}) = - \sin[\theta-\theta^*(\bm{r})]~.
\end{equation}
Namely, we assume that the particle seeks to align with a stationary vector field $\hat{\bm{u}}^*(\bm{r}) = (\cos\theta^*(\bm{r}),\sin\theta^*(\bm{r}))$, which we will henceforth refer to as the \emph{steering policy} \cite{piro2022efficiency,piro2022optimal}. 
In physical terms, the particle exerts feedback control on its heading direction based on local measurements of its current position, following a predefined steering strategy. 
This setup is illustrated schematically in Fig.~\ref{fig:intro_schematic}. 
The origin of the policy itself, whether provided externally or established autonomously, e.g., via reinforcement learning~\cite{stark2021artificial,schneider2019optimal,muinos2021reinforcement,biferale2019zermelo,putzke2023optimal,Nasiri_2023}, is immaterial for the problem at hand. 

Rescaling space and time as $\bm{r} \to (w/D_\theta)\bm{r}$ and $t \to t/D_\theta$, and introducing the dimensionless P{\'e}clet number ${\rm Pe} \equiv w^2/(D_t D_\theta)$ and alignment strength $\kapt \equiv \kappa/D_\theta$, 
the probability density $P(\bm r,\theta,t)$ associated with Eqs.~(\ref{eq:lang_r},\ref{eq:lang_t}) obeys the Fokker-Planck equation
\begin{equation}\label{eq:fp_full}
    \partial_t P(\bm{r},\theta,t) = -\tilde{\nabla} \cdot  \bm{J},
\end{equation}
with $\tilde{\nabla} \equiv (\partial_x,\partial_y,\partial_\theta)$ and the probability current 
\begin{eqnarray}
\bm{J} = 
    \left(\begin{array}{c}
           J_x \\
           J_y \\
           J_\theta
    \end{array}\right) = 
    \left(\begin{array}{c}
           (\cos\theta + v_x) P - {\rm Pe}^{-1} \partial_x P \\
           (\sin\theta + v_y) P - {\rm Pe}^{-1} \partial_y P \\
           -\kapt \sin(\theta-\theta^*)P - \partial_\theta P
    \end{array}\right)~.
\end{eqnarray}

\paragraph{Entropy Production --}
We equip the dynamics in Eq.~\eref{eq:fp_full} with a thermodynamic interpretation by computing the mean rate of entropy production at steady state \cite{seifert2012stochastic,cocconi2020entropy}. This is proportional to the mean stochastic work done by the active forces \cite{speck2011work,lee2023nonequilibrium}, and can be written as
\begin{eqnarray}
    \dot{S} 
    &= \int \rmd\theta \rmd\bm{r} \ {\rm Pe}[(J_x - v_x P_{\rm s})\cos\theta + (J_y - v_y P_{\rm s})\sin\theta] - \kapt\sin(\theta-\theta^*)J_\theta \nonumber \\
    &= {\rm Pe} + \kapt[\kapt\langle \sin^2(\theta-\theta^*)\rangle - \langle \cos(\theta-\theta^*) \rangle]~, \label{eq:epr_exact}
\end{eqnarray}
where $\langle \cdot \rangle$ denotes the average with respect to the steady-state distribution $P_{\rm s}(\bm{r},\theta)$,
while the second line was obtained via integration by parts and after discarding boundary terms.
Note that~\eref{eq:epr_exact} is valid for any choice of steering policy $\hat{\bm{u}}^*$\footnote{This result can equivalently be derived by writing the entropy production rate as the Kullback-Leibler divergence per unit time of the ensemble of forward $(\bm{r}, \theta)$ trajectories and
their time-reversed counterparts \cite{gaspard2004time,cocconi2020entropy}, taking care to reverse the direction of the external flows in the time-reversed ensemble so as to enforce Galilean invariance \cite{Speck2008}.}.

We identify the two terms in the right-hand side of Eq.~\eref{eq:epr_exact} as distinct contributions to the dissipation originating from the self-propulsion ($\dot{S}_{\rm p} = {\rm Pe}$) and feedback control on the heading angle 
\begin{equation}\label{eq:dots_c}
    \dot{S}_{\rm c} = \kapt[\kapt\langle \sin^2(\theta-\theta^*)\rangle - \langle \cos(\theta-\theta^*) \rangle]~,
\end{equation}
respectively.
Importantly, only the second contribution depends on the policy $\hat{\bm{u}}^*$.
Below, we compare our theoretical predictions with direct simulations of the Langevin equations~(\ref{eq:lang_r},\ref{eq:lang_t}) by evaluating $\dot{S}_{\rm c}$
in the Sekimoto framework \cite{sekimoto2010stochastic} from the time-averaged stochastic work computed over steady-state trajectories. 

\subsection{The gradient expansion approximation}\label{ss:moment_exp_brief}

Although Eq.~\eref{eq:dots_c} is exact, evaluating $\dot{S}_{\rm c}$ in practice requires the knowledge of the steady state solution of Eq.~\eref{eq:fp_full},
which is, in general, not known exactly.
Here, we therefore introduce an approximation scheme
based on a gradient expansion of $P(\bm r,\theta,t)$,
that will enable us to evaluate the steady state moments appearing in Eq.~\eref{eq:dots_c} when the spatial variations of $P$ are sufficiently smooth.
Below, we only sketch the key steps and report the main results, with the complete derivation outlined in~\ref{a:mom_exp_detail}. 

Assuming, in the spirit of an adiabatic approximation, that the relaxation of the angular degree of freedom $\theta$ is fast compared to  that of the swimmer position, we seek factorized solutions of Eq.~\eref{eq:fp_full} of the form
\begin{equation}\label{eq:P_exp_mom}
    P(\bm{r},\theta,t) = \rho(\bm{r},t)\mathcal{Q}(\theta,\bm{r}) = \rho(\bm{r},t) [\mathcal{Q}_0(\theta,\bm{r}) + \mathcal{Q}_1(\theta,\bm{r}) + ...] ,
\end{equation}
where $\rho$ denotes the marginal probability density,
while the orientational part of the distribution is normalized as $\int_0^{2\pi}\rmd\theta \calQ(\theta,\bm r) = 1$ and the second equality has been obtained by expanding $\calQ$ into contributions
$\mathcal{Q}_n = \mathcal{O}(\nabla^n)$ of increasing order in the spatial gradient. 
The dynamics of $\rho$ is obtained after integrating \eref{eq:fp_full} with respect to the orientation $\theta$, giving
\begin{equation}\label{eq:rho_dyn}
    \partial_t \rho + \nabla\cdot \left[ \left( \langle \hu \rangle_{\theta} + \bm v - \Pe^{-1}\nabla \right) \rho \right] = 0,
\end{equation}
and where we have defined $\langle \cdot \rangle_{\theta} \equiv \rho^{-1}\int_0^{2\pi}\rmd\theta \, (\cdot) P = \int_0^{2\pi}\rmd\theta \, (\cdot) \mathcal{Q}$. 
At this level of coarse-graining, the self-propulsion simply renormalizes the advection term.
As detailed in~\ref{a:mom_exp_detail}, the equations satisfied by the $\calQ_n$ contributions are obtained by inserting the ansatz~\eref{eq:P_exp_mom} in~\eref{eq:fp_full} and matching terms of the same order in $\nabla$.
This way, we obtain an infinite hierarchy where the equation for the distribution $\calQ_n$ only depends on the lower order contributions $\calQ_k$ with $k \in [0;n-1]$.
In principle, the hierarchy can be solved iteratively at arbitrary order in $\nabla$.
In practice, the expressions of the distributions $\calQ_n$ quickly become lengthy and provide only limited insights.
For practical purposes, we thus limit the derivation to the first nontrivial order $n=1$, which allows us to determine the effective drift and diffusivity that arise from the elimination of the orientation.

The $n=0$ order corresponds to a strict adiabatic approximation of the orientational degrees of freedom, as it amounts to setting $J_\theta = 0$.
The corresponding solution is therefore
\begin{equation} \label{eq_Q0_def}
    \mathcal{Q}_0(\theta,\bm{r}) = \frac{e^{\kapt \cos[\theta-\theta^*(\bm{r})]}}{2\pi I_0(\kapt)}~,
\end{equation}
where $I_k$ is the modified Bessel function of the first kind of order $k$.
To obtain tractable expressions at ${\cal O}(\nabla)$, we moreover work at linear order in $\kapt$.
After some algebra detailed in~\ref{a:mom_exp_detail}, we obtain
\begin{eqnarray}
\frac{\calQ_1(\theta,\bm r)}{\calQ_0(\theta,\bm r)} = &-\left[\left( \hu(\theta) - \frac{\kapt}{2}\hu^* - \frac{\kapt}{8}\hu(2\theta - \theta^*) \right)\cdot\nabla\right]\ln\rho \nonumber \\
&- \kapt \left[ 
(\bm v \cdot \nabla)(\hu(\theta) \cdot \hu^*)
+ \nabla \cdot \left( \hu(2\theta - \theta^*) \right)
\right]
+ {\cal O}(\kapt^2)~.
\end{eqnarray}

Using $\calQ_0$ and $\calQ_1$ to evaluate the average orientation in Eq.~\eref{eq:rho_dyn},
the density field effectively obeys a drift diffusion equation
\begin{equation}\label{eq:cg_fp_rho_main}
\partial_t \rho + \nabla\cdot \left[ \left( \bm v_{\rm eff} - D_{\rm eff}\nabla \right) \rho \right] = 0,
\end{equation}
with effective parameters
\begin{equation*}
\bm v_{\rm eff} = \bm v + \frac{\kapt}{2} (1 - \bm v\cdot\nabla)\hu^*, \qquad
D_{\rm eff} = \Pe^{-1} + \frac{1}{2}.
\end{equation*}
After evaluating the moments involved in the right-hand side of Eq.~\eref{eq:dots_c}, we find that it simplifies to
\begin{equation}\label{eq:sdot_c_11mom}
\dot{S}_{\rm c} 
= -\frac{\kapt}{2}\left\langle \nabla \cdot \hu^* \right\rangle + {\cal O}(\kapt^2,\nabla^2)~,
\end{equation}
with $\langle \cdot \rangle$ now denoting averaging with respect to $\rho_{\rm s}(\bm{r})$ as obtained by solving Eq.~\eref{eq:cg_fp_rho_main} in the steady state. 
The expression~\eref{eq:sdot_c_11mom} for the rate of entropy production resulting from feedback control is one of the key results of this work.
In contrast with Eq.~\eref{eq:dots_c}, it can be used to estimate the corresponding dissipation directly from measurements of the steady state density $\rho_s$,
and without the knowledge of the full joint distribution  $P(\bm{r},\theta)$.

Equation~\eref{eq:sdot_c_11mom} further formalizes the intuition that policies aiming to localise the swimmer in the proximity of a target subspace inevitably incur a thermodynamic cost since density accumulates in regions of negative policy divergence (Fig.~\ref{fig:intro_schematic}). 
Indeed, a cost-free policy (i.e., for which $\dot{S}_{\rm c} = 0$) requires the existence of a stream-like function $\Psi(\bm r)$ such that 
$\hu^* = (\partial_y \Psi,-\partial_x \Psi)$.
Imposing that $\hu^*$ has a unit norm everywhere, the only smooth vector field satisfying this constraint in $\mathbb{R}^2$ corresponds to a 
uniform drift, which does not require adaptation.
More generally, any divergence-free policy cannot localise particles since, from the divergence theorem, it must induce a density current satisfying a zero net flux condition through any closed surface.

At the level of approximation of Eq.~\eref{eq:sdot_c_11mom}, 
the advection $\bm{v}$ does not appear explicitly in the expression of $\dot{S}_{\rm c}$.
We show in~\ref{a:mom_exp_detail} that this feature holds at all orders in $\kapt$, while additional contributions involving $\bm v$ 
will arise when the expansion is carried out at ${\cal O}(\nabla^2)$ (see~\ref{app_Onabla2_exp} for a detailed discussion).
However, we note that even at ${\cal O}(\nabla)$, the effect of $\bm v$ features implicitly both via the steady state density $\rho_{\rm s}$, as well as potentially through $\hat{\bm{u}}^*$, which would in general depend on the local flow landscape~\cite{piro2022efficiency,piro2022optimal}. 

\section{Localisation at a point target}\label{s:pt_target}

Having established the general framework for the thermodynamic characterisation of a self-steering motile particle, 
we now move on to the study of specific policies. 
We start with perhaps the simplest relevant setup, namely the steering policy $\hat{\bm{u}}^*(\bm{r}) = - \hat{\bm{r}}$ designed to localise the agent at a point target located 
at $\bm{r}^*= \bm 0$ on an infinite plane. 
To calculate the dissipation associated with feedback control, $\dot{S}_{\rm c}  = \frac{\kapt}{2}\langle 1/r \rangle$ with $r=|\bm{r}|$, 
we need to determine the steady state solution of Eq.~\eref{eq:cg_fp_rho_main}, which depends on the form of the imposed flow $\bm v(\bm r)$. 

\paragraph{Localisation without flow --} 
First, we consider the case where the dynamics is not subject to an externally applied flow, $\bm{v} =\bm 0$. 
The steady-state probability density solution of Eq.~\eref{eq:cg_fp_rho_main} is then exponentially localised around the target:
\begin{equation}\label{eq:rho_1pt_noflow}
    \rho_{\rm s}(\bm{r}) = \frac{1}{2\pi\ell^2}e^{-r/\ell},\qquad \ell \equiv \frac{2 D_{\rm eff}}{\tilde{\kappa}}~.
\end{equation}
To measure the accuracy of the policy, we introduce the steady state variance of the radial displacement: 
$\sigma_r^2 \equiv \int \rmd \bm{r} \, r^2 \rho_{\rm s}(\bm{r})$, 
which for the solution~\eref{eq:rho_1pt_noflow} is equal to $6 \ell^2$.
On the other hand, the rate of entropy production $\dot{S}_{\rm c}$ quantifies energy expenditure associated with feedback control, 
and  is given via Eqs.~(\ref{eq:sdot_c_11mom},\ref{eq:rho_1pt_noflow}) as $\dot{S}_c = D_{\rm eff}/\ell^2$. 
Taken together, these two measures of performance thus entail
\begin{equation}\label{eq:tradeoff_1pt_noflow}
    \sigma_r^2 \dot{S}_c = 6 D_{\rm eff}~.
\end{equation}
The equality~\eref{eq:tradeoff_1pt_noflow} clearly highlights that we are faced here 
with a multi-objective optimization problem of balancing precision and energy expenditure, i.e.\ with a dissipation-accuracy tradeoff. 
In particular, Eq.~\eref{eq:tradeoff_1pt_noflow} predicts a divergence of $\dot{S}_{\rm c}$ as $\sigma_r \to 0$.
This feature results from the discontinuous nature of the aligning policy at the point target,
which physically translates into an increase in the frequency of reorientations with the accuracy of the localisation.
Conversely, for large $\sigma_r$ the particle spends most of the time far away from the target where $\hu$ is nearly uniform, and thus only experiences weak reorientations.
Although we have derived Eq.~\eref{eq:tradeoff_1pt_noflow} for the simplest problem of localisation, 
we will show below that it bears a certain degree of generality, as similar relations are obtained for more complex cases, including 
nonzero applied flow and localisation along a one-dimensional path.

\paragraph{Localisation with a radial flow --} 
A straightforward generalisation of the previous example considers the influence of a radially symmetric stationary flow field.
Assuming for simplicity that the vorticity vanishes everywhere, $\nabla \times \bm{v}=\bm 0$, we
write $\bm v = -V'(r) \hat{\bm{r}}$ in terms of an effective potential $V$. As such, the flow need not be divergence-free, 
which could be achieved, e.g., by the presence of distributed sources/sinks coupled with a third dimension. 
From the expression of the resulting effective radial drift $\bm v_{\rm eff} = -[V'(r) + \kapt/2]\hat{\bm r}$, 
we require $V'(r) > -\kapt/2$, i.e. that the renormalised self-propulsion speed exceeds the flow velocity, such that Eq.~\eref{eq:cg_fp_rho_main} admits a nonvanishing stationary solution:
\begin{equation} \label{eq_density_radial_confinement}
\rho_{\rm s}(r) = \mathcal{N} \exp\left[-\frac{r}{\ell} - \tilde{V}\left(\frac{r}{\ell_v}\right) \right].
\end{equation}
Here, we further assumed that the flow amplitude is essentially controlled by a single length scale $\ell_v$, 
so that we have defined $\tilde{V}(r/\ell_v) = V(r)/D_{\rm eff}$, and $\mathcal{N}$ ensures the normalization of $\rho_{\rm s}$.
For example, in the case $V(r) = v_{\rm f} r$ of a linear potential, $\ell_v = D_{\rm eff} / |v_{\rm f}|$, such that the weak and strong flow regimes correspond to large and small $\ell_v$, respectively.

Computing the measures of accuracy and thermodynamic efficiency,   
we find that they take the general expressions
\begin{equation} \label{eq_sigma_epr_radial_potential}
\frac{\sigma_r^2}{\ell^2} = \frac{{\cal I}_3\left( \frac{\ell}{\ell_v} \right)}{{\cal I}_1\left( \frac{\ell}{\ell_v} \right)}~, \qquad
\frac{\ell^2 \dot{S}_c}{D_{\rm eff}} = \frac{{\cal I}_0\left( \frac{\ell}{\ell_v} \right)}{{\cal I}_1\left( \frac{\ell}{\ell_v} \right)}~,
\end{equation}
where ${\cal I}_n(x) = \int_0^\infty \rmd y\, y^n e^{-y-\tilde{V}(x y)}$.
In their rescaled forms~\eref{eq_sigma_epr_radial_potential}, both the variance of radial displacement and the rate of entropy production are uniquely controlled by the ratio of the length scales associated with feedback control and the externally applied flow, respectively $\ell$ and $\ell_v$.
In the limit $\ell \ll \ell_v$ of weak flow,
the integrals ${\cal I}_n(\ell/\ell_v) \simeq \Gamma(n+1)$ for $n \ge 0$ with $\Gamma$ denoting the Gamma function, such that we naturally recover the results obtained in the previous section.

\begin{figure}
    \centering
    \includegraphics[width=\columnwidth]{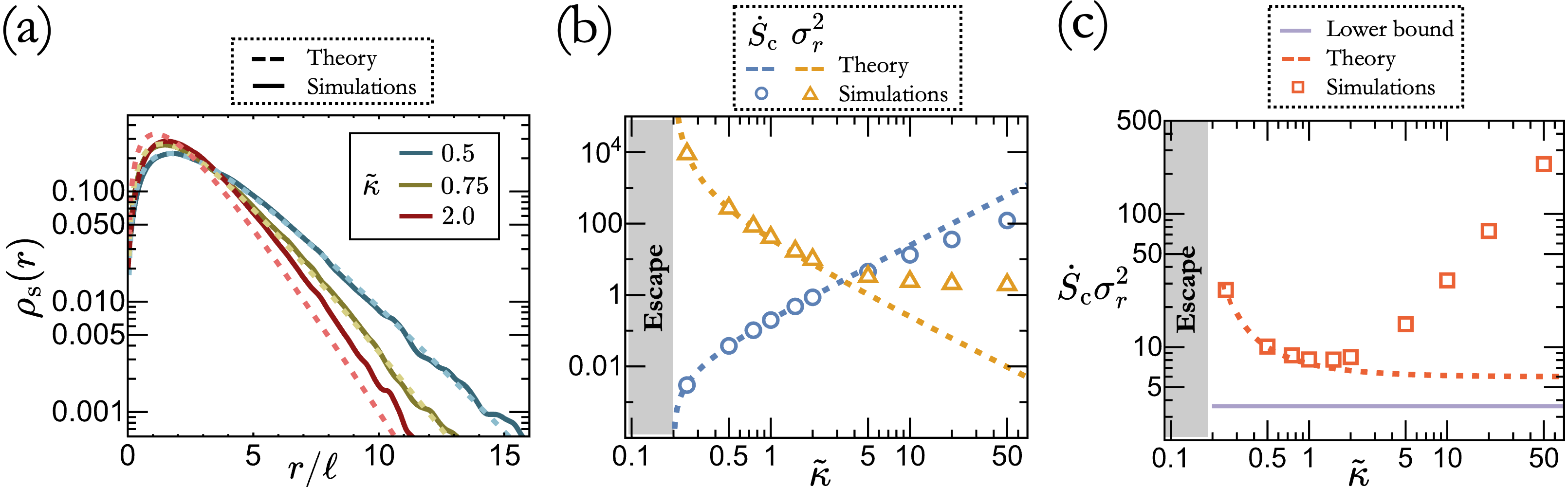}
    \caption{Localisation at a point target in the presence of a divergent flow field.
    (a) Radial steady-state density distributions for a particle self-localising around a target at $r=0$. Dashed lines correspond to the theoretical prediction Eq.~(\ref{eq_density_radial_confinement}).
    (b) Entropy production rate ($\dot{S}_{\rm c}$) and variance of the radial displacement ($\sigma_r^2$) as functions of the dimensionless alignment strength $\kapt$. Theoretical predictions are given in Eq.~(\ref{eq_sigma_epr_radial_potential_linearV}).
    (c) $\dot{S}_{\rm c}\sigma_r^2$ as a function of $\kapt$. The red and purple colors correspond to the cases with (Eq.~(\ref{eq_sigma_epr_radial_potential_linearV})) and without (Eq.~(\ref{eq:tradeoff_1pt_noflow})) divergent flow, respectively.
    In all panels $\Pe = 2$ and $\ell_v = 10$.}
    \label{fig:point_target}
\end{figure}

On the other hand, a strongly convergent flow ($V'(r)>0$) can localise the particle without relying on feedback control, naturally leading to the disappearance of the dissipation-accuracy tradeoff.
Here, we thus rather examine the scenario of a divergent flow ($V'(r) < 0$) competing with the steering
policy by advecting the particle away from the target.
We assume for simplicity that the potential is linear: $\tilde{V}(r/\ell_v) = -r/\ell_v$, ---whence the constant flow strength $v_{\rm f} = D_{\rm eff}/\ell_v$--- such that the steady-state density~\eref{eq_density_radial_confinement} remains exponential with a characteristic length scale $(\ell^{-1} - \ell_v^{-1})^{-1}$ (Fig.~\ref{fig:point_target}(a)). 
localisation at a finite distance to the target therefore implies $\ell / \ell_v < 1$, 
while $\ell = \ell_v$ leads to the escape of the particle.
In dimensional units, the condition $\ell < \ell_v$ corresponds to the lower bound $w > 2 D_\theta v_{\rm f}/\kappa$ on the bare self-propulsion speed.
Hence, we again find that in the regime of weak alignment strength considered here, the condition $w > v_{\rm f}$ is not sufficient to ensure that the agent remains in close vicinity of the target.

To evaluate the thermodynamic cost associated with localisation, we calculate the integrals in Eq.~\eref{eq_sigma_epr_radial_potential} explicitly, and obtain 
\begin{equation} \label{eq_sigma_epr_radial_potential_linearV}
    \sigma_r^2 = \frac{6\ell^2}{(1 - \ell/\ell_v)^2}, \qquad
    \dot{S}_{\rm c} = \frac{D_{\rm eff}}{\ell^2}\left( 1 - \frac{\ell}{\ell_v}\right).
\end{equation}
As shown in Fig.~\ref{fig:point_target}(b), the particle escape is simultaneously associated with the cancellation of $\dot{S}_{\rm c}$ and the divergence of $\sigma_r^2$.
Furthermore, the product $\sigma_r^2\dot{S}_{\rm c} = 6 D_{\rm eff}/(1-\ell/\ell_v)$ is always bounded from below by the right hand side of~\eref{eq:tradeoff_1pt_noflow} (Fig.~\ref{fig:point_target}(c)).
This is in agreement with the intuition that
actively localising the particle at the target requires more dissipation in the presence of a divergent flow.
In particular, tuning $\ell$ towards $\ell_v$, i.e., by weakening the active localisation, the entropy production rate required to maintain a fixed variance $\sigma^2_r$ can be made arbitrarily large.
In the opposite limit of strong active localisation ($\ell \ll \ell_v$), although Eq.~\eref{eq_sigma_epr_radial_potential_linearV} predicts that $\sigma_r^2\dot{S}_{\rm c}$ converges to $6 D_{\rm eff}$, the numerical results presented in Fig.~\ref{fig:point_target}(c) rather show that the product grows with $\kapt$. 
We interpret this discrepancy as the breakdown of the expansion performed in Sec.~\ref{ss:moment_exp_brief}, which formally assumes low $\kapt$, and fails to capture the saturation of the variance $\sigma^2_r$ at large $\kapt$ (see Fig.~\ref{fig:point_target}(b)).
The monotonous increase of $\sigma_r^2\dot{S}_{\rm c}$ at $\kapt \gtrsim 5$ observed in numerical simulations then reflects the fact that the dissipation associated with feedback control always increases with the alignment strength (see Eq.~\eref{eq:dots_c}).

As shown in~\ref{a:target_uniform} for the case of localisation in a uniform flow field, 
the phenomenology described here does not qualitatively depend on the form or symmetry of the applied flow, so long as the escape occurs below a finite threshold of alignment strength.  

\section{Motion along a target path}\label{s:target_path}

We now move on to the study of steering policies $\hat{\bm{u}}^*$ designed to localise the self-steering motile particle along a directed one-dimensional subspace, which we refer to as a target path. 
This setup is particularly relevant in the context of stochastic optimal navigation, where strategies optimizing travel time or energy may involve the swimmer following predetermined target trajectories~\cite{schneider2019optimal,muinos2021reinforcement,piro2022optimal,piro2024energetic}.
The ability of ``smart'' agents to traverse complex environments along reproducible paths in the presence of noise is equally essential in the context of autonomous shape assembly~\cite{sun2023mean,rubenstein2014programmable,slavkov2018morphogenesis}, foraging \cite{CzaczkesAnnRev2015}, exploration and patrolling \cite{YangSciRob2018}, as well as targeted delivery \cite{LiSciRob2017}. 

%
For simplicity, we take the target path to be a straight line coinciding with the $x$ coordinate axis, $y=0$. 
This approximation shall be well verified so long as the typical radius of curvature of the target path is well separated from the characteristic lengthscale associated with the localisation.
Below, we consider policies which essentially aim at stabilizing the position of the agent near the target path via an aligning term that prioritises orthogonal (longitudinal) motion at large (small) orthogonal displacements (see Fig.~\ref{fig:intro_schematic}(b) for a schematic example).
In the presence of transverse flows that deviate the active particle from the target path, 
we further show how these policies can be augmented with an additional term compensating for the local transverse drift, 
and which we refer to as \emph{compensating} policies. 
%

\subsection{Localisation without flow}\label{s:purely_align}

In this section, we set $\bm{v}(\bm{r}) = \bm{0}$ so that the agent does not experience advection by an external flow. 
The steering policy $\hat{\bm{u}}^*(y)$ then depends only on the (signed) orthogonal displacement from the target path. 

\subsubsection{Probability density of the orthogonal displacement}

By construction, the problem is translationally invariant with respect to the longitudinal coordinate $x$.
We thus integrate Eq.~\eref{eq:cg_fp_rho_main} over $x$ and obtain the Fokker-Planck equation for the 
marginal density distribution associated with orthogonal displacement
\begin{equation}\label{eq:fp_ort_comp}
    \partial_t \varrho(y,t) = D_{\rm eff} \partial_y \left[ \partial_y \varrho(y,t) - \frac{\varrho(y,t) \sin\theta^*(y)}{\ell} \right],
\end{equation}
where $\varrho(y,t) \equiv \int\rmd x\, \rho(x,y,t)$, and $\ell = 2 D_{\rm eff}/\kapt$ as defined in the previous section. 
Equation~\eref{eq:fp_ort_comp} describes the one-dimensional dynamics of a Brownian particle with diffusivity $D_{\rm eff}$
in a  pseudo-potential $D_{\rm eff} U(y)/\ell$, with $U'(y) = -\sin\theta^*(y)$. 
The steady-state solution for this equilibrium problem is thus given by the Boltzmann measure
\begin{equation}\label{eq:rho_y_analyt}
    \varrho_{\rm s}(y) = \mathcal{N} {\rm exp}\left( - \frac{U(y)}{\ell} \right),
\end{equation}
with $\mathcal{N}^{-1} \equiv \int\rmd y\, \varrho_{\rm s}(y)$ a normalisation factor. 
Drawing on Eq.~\eref{eq:rho_y_analyt} for a given choice of $U(y)$, one can now compute observables of physical interest, such as the steady-state variance of the distance from the target path
\begin{equation}\label{eq:var_momexp}
    \sigma_y^2 =  \int \rmd y \, \varrho_{\rm s}(y) y^2~,
\end{equation}
as well as the mean rate  of entropy production associated with steering control, via Eq.~\eref{eq:sdot_c_11mom},
\begin{equation}\label{eq:epr_momexp}
    \dot{S}_{\rm c} = \frac{D_{\rm eff}}{\ell} \int \rmd y \, \varrho_{\rm s}(y) U''(y)~.
\end{equation}

\subsubsection{The adaptive aligning policy} \label{ss:aap}


A straightforward generalization of the policy introduced in Sec.~\ref{s:pt_target} corresponds to $\sin\theta^*(y) = -y/|y|$, 
i.e. imposing the agent to orient its direction of motion orthogonally to the target path.
From a navigation perspective, however, a more efficient strategy would rather maximize the displacement along the target path when the swimmer is sufficiently close to it.
Hence, below we consider 
a one-parameter family of aligning policies, defined by 
\begin{equation}\label{eq:lin_pol}
    \sin\theta^*(y) = - \frac{y}{{\rm max}(\varepsilon,|y|)} , \qquad
    \cos\theta^*(y) = \sqrt{1 - \sin^2\theta^*},
\end{equation}
the sign of the cosine being chosen to induce a drift in the positive $x$ direction. 
Equation~\eref{eq:lin_pol} essentially corresponds to the ``adaptive" aligning policy (AAP) introduced in Ref.~\cite{piro2022optimal}.
It features an additional parameter, $\varepsilon \ge 0$, which controls the characteristic length scale of localisation independently of the alignment strength $\kapt$. 
As we show below, in addition to improving navigation along the target path, the presence of $\varepsilon > 0$ 
also helps to achieve a better dissipation-accuracy tradeoff for localisation.

\begin{figure}[t!]
    \centering
    \includegraphics[width=\columnwidth]{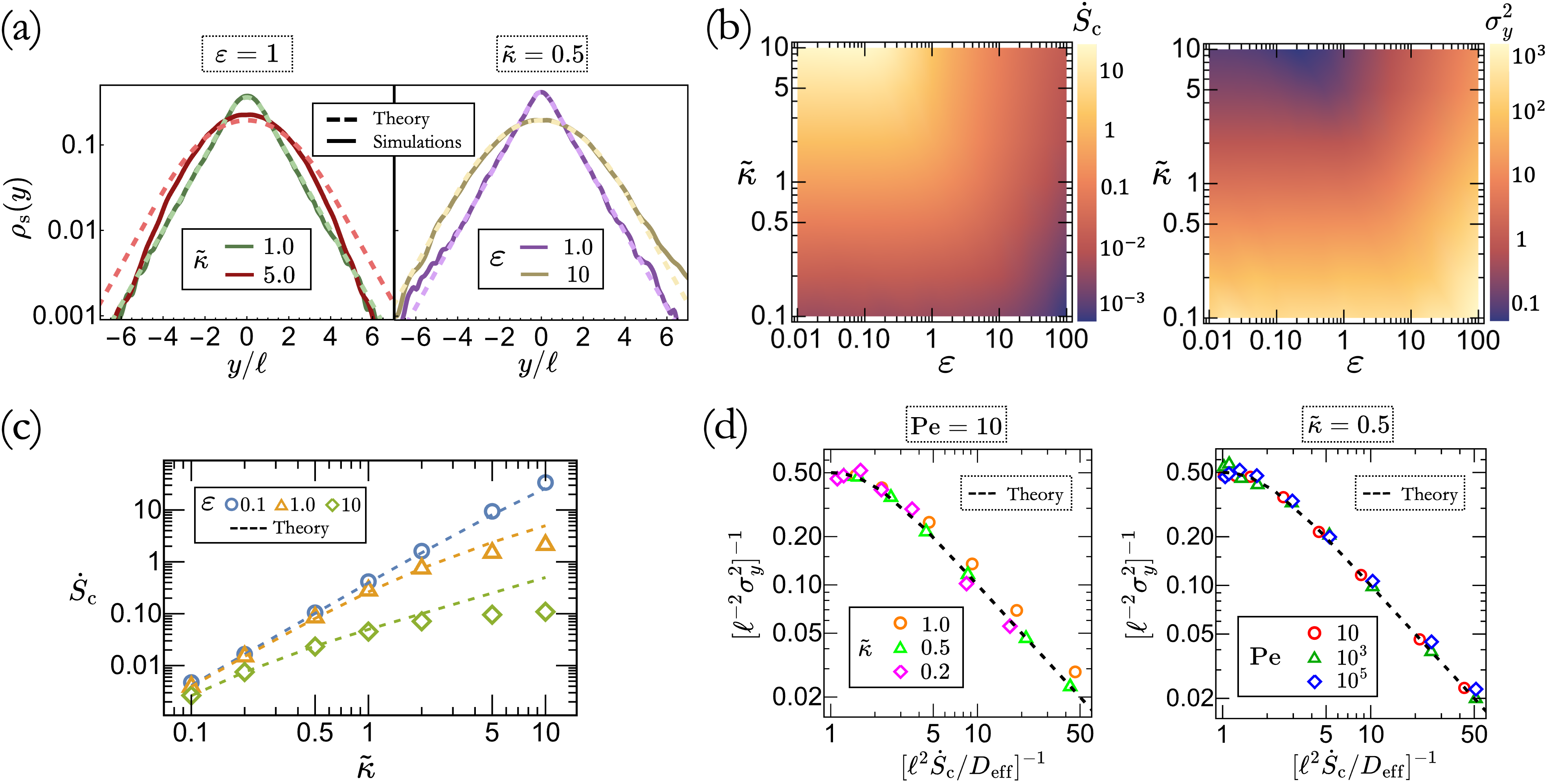}
    \caption{Comparison between theory and simulations of the adaptive aligning policy (AAP) described in Sec.~\ref{ss:aap}. (a) Probability density functions of the orthogonal displacement obtained from the numerical simulations (solid lines) compared to those predicted by Eq.~(\ref{eq:rho_cases}) (dashed lines) for different values of $\kapt$ and $\varepsilon$. 
    (b) Color maps showing the steady-state entropy production rate and variance of the orthogonal displacement computed from direct simulations of the Langevin dynamics as a function of the alignment strength $\kapt$ and confinement scale $\varepsilon$. 
    (c) Mean entropy production rate obtained from simulations (open markers) and from Eq.~(\ref{eq:O_nabla_kappa_epr}) (dashed lines), shown as a function of $\kapt$. 
    (d) Data collapse of the rescaled entropy production and variance (cf.~Eqs.~(\ref{eq:O_nabla_kappa_sigma},\ref{eq:O_nabla_kappa_epr})) for different values of $\kapt \lesssim1$ and $\Pe \gtrsim 10$ as indicated by the legend, setting $\varepsilon=1$ in both cases. 
    }
    \label{fig:AAP}
\end{figure}

Substituting Eq.~\eref{eq:lin_pol} in the definition of the pseudo-potential $U(y)$, we obtain the 
piecewise-defined steady-state probability density
\begin{equation} 
\varrho_{\rm s}(y) = \frac{1}{2\ell}{\cal F}\left(\frac{\varepsilon}{2\ell}\right)\times
\left\{\begin{array}{ll}
    e^{(\varepsilon - y)/\ell} & y>\varepsilon \\
    e^{(\varepsilon^2-y^2)/(2 \varepsilon \ell)} & |y| < \varepsilon \\
    e^{(\varepsilon + y)/\ell} & y<-\varepsilon 
\end{array}\right. ,
\quad \ell = \frac{2 D_{\rm eff}}{\kapt}~,
\label{eq:rho_cases}
\end{equation}
with ${\cal F}(x) = 1/[1 + \sqrt{\pi x } {\rm Erf}\left(\sqrt{x}\right) e^x]$.
The density $\varrho_{\rm s}$ exhibits a Gaussian bulk for $|y| < \varepsilon$, followed by exponential tails for $|y| > \varepsilon$. 
For $\kapt \lesssim 1$, the prediction~\eref{eq:rho_cases} is in excellent agreement with empirical histograms constructed from numerical simulations of the Langevin dynamics~(\ref{eq:lang_r},\ref{eq:lang_t}), as shown in Fig.~\ref{fig:AAP}(a). 

\mycomment{We notice that Eqs.~\eref{eq:rho_cases} simplifies, for ${\rm Pe} \gg 1$, to
\begin{equation}
    \varrho(y) \simeq \frac{\tilde\kappa}{\sqrt{2\pi \varepsilon\tilde\kappa} \ {\rm exp}\left( \frac{\varepsilon\tilde\kappa}{2} \right) {\rm Erf}\left( \sqrt{\frac{\varepsilon\tilde\kappa}{2}}\right)  }
\end{equation}
which is independent of Pe itself.}

Furthermore, Eq.~\eref{eq:rho_cases} allows to obtain explicit estimates for the observables measuring the accuracy of the policy and the thermodynamic cost associated to it.
Namely, using Eqs.~(\ref{eq:var_momexp}-\ref{eq:rho_cases}) and after some algebra, we obtain
\begin{eqnarray}
\label{eq:O_nabla_kappa_sigma}
\frac{\sigma_y^2}{2\ell^2} & = {\cal F}\left(\frac{\varepsilon}{2\ell}\right) + \frac{\varepsilon}{2\ell}\left[1 + {\cal F}\left(\frac{\varepsilon}{2\ell}\right)\right]~, \\
\label{eq:O_nabla_kappa_epr}
\frac{\ell^2 \dot{S}_{\rm c}}{D_{\rm eff}} & = \frac{\ell}{\varepsilon}\left[1 - {\cal F}\left(\frac{\varepsilon}{2\ell}\right)\right]~.
\end{eqnarray}
Studying the variations of the function ${\cal F}$, we find that the dissipation and accuracy satisfy the following inequalities:
\begin{equation} \label{eq:ineq_prod}
\frac{1}{2} \le \frac{\dot{S}_{\rm c} \sigma_y^2}{2 D_{\rm eff}} \le 1~,
\end{equation}
where the upper and lower bounds are met in the limits $\varepsilon = 0$ and $\varepsilon/\ell \to \infty$, respectively.
As for the point target, Eq.~\eref{eq:ineq_prod} highlights the dissipation-accuracy tradeoff associated with localisation along a target path. 
Since the bounds in Eq.~\eref{eq:ineq_prod} correspond to the regimes of weak and strong confinement along the target path, 
we speculate that they also hold for a broader class of pseudo-potentials.
We support this conjecture with numerical simulations in the following section. 

This tradeoff is also clearly exemplified by the opposite trends, in both the alignment strength $\kapt = D_{\rm eff}/\ell$ and $\varepsilon$,
observed when comparing mean rate of entropy production and variance obtained from direct numerical simulations of the Langevin dynamics~(\ref{eq:lang_r},\ref{eq:lang_t}), as shown in Fig.~\ref{fig:AAP}(b).
Moreover, we show in Fig.~\ref{fig:AAP}(c) that the dependence of $\dot{S}_{\rm c}$ on $\kapt$ and $\varepsilon$ is well captured at small $\kapt$ by our $\mathcal{O}(\kapt,\nabla)$ result, Eq.~\eref{eq:O_nabla_kappa_epr}.

We further confirm in Fig.~\ref{fig:AAP}(d) that by rescaling $\sigma_y^2$ and $\dot{S}_{\rm c}$ with appropriate powers of $\ell$ and $D_{\rm eff}$, the data representing dissipation as a function of precision collapse onto a master curve following, for large $\varepsilon/\ell$, a power law relationship with negative exponent $-1$. 
On the other hand, for small values of $\varepsilon/\ell$, the curves in Fig.~\ref{fig:AAP}(d) are concave, which can be quantified by expanding Eqs.~(\ref{eq:O_nabla_kappa_sigma},\ref{eq:O_nabla_kappa_epr}) around $\varepsilon = 0$, giving $\sigma^2/2\ell^2 \simeq 1 + \frac{2}{3}(\varepsilon/\ell)^2$ and $\ell^2 \dot{S}_{\rm c}/D_{\rm eff} \simeq 1 - 2\varepsilon/\ell$.
In practice, this indicates that aligning the moving direction of the swimmer with the target path in its vicinity generally leads to reduced dissipation without significant loss of accuracy.

To conclude this section, we note that the above findings are qualitatively similar to results obtained from an exactly solvable version of the model~(\ref{eq:lang_r},\ref{eq:lang_t}) where the swimmer's heading direction can only take discrete values: $\theta = \pm\phi$ with $\phi \leq \pi/2$.
In particular, as detailed in~\ref{app:aligning_policy}, the scaling behavior $\dot{S}_{\rm c} \simeq \sigma_y^{-2}$ obtained in the regime of weak localisation is also recovered for this case (cf.~Eq.~\eref{eq:large_eps_scaling} and Fig.~\ref{fig:ap_curves}).

\subsubsection{Optimal localisation policy}
\label{sec_optimal_policy}
Although the Boltzmann-like distribution~\eref{eq:rho_y_analyt} formally approximates the transverse displacement statistics only for $\kapt \lesssim 1$ and weakly curved paths, it nevertheless allows to efficiently characterize a large ensemble of policies without having to determine the full probability density $P(\bm r,\theta,t)$.
In this section, we extend the approach and show how the framework presented above can be used in practice to determine a set of optimal policies. 

To proceed, we parametrize the pseudo-potential $U(y)$ demanding on the basis of symmetry and intuition that it must remain an even, convex function.
Additionally, its derivative needs to be interpretable according to $U'(y) = -\sin\theta^*$, such that $|U'(y)| \leq 1$ for all $y \in \mathbb{R}$. 
We therefore choose a polynomial parametrization by defining 
\begin{equation}\label{eq:def_tildeU}
    \tilde{U}(y) = \sum_{n=1}^M a_n y^{2n}~, \qquad y^* = {\rm min}\{ y \; ; \; \tilde{U}'(y) = 1\}~,
\end{equation}
where all $a_n \ge 0$.
The pseudo-potential is then defined as a piecewise function,
\begin{equation}
    U(y) = 
    \left\{ \begin{array}{ll}
    -( y + y^* ) + \tilde{U}^* & y<-y^* \\
    \tilde{U}(y) & |y| \le y^* \\
     y - y^* + \tilde{U}^* & y> y^* 
\end{array}\right.~,
\end{equation}
where $\tilde{U}^* = \tilde{U}(y^*)$ for continuity.
The adaptive aligning policy discussed in the previous section is recovered by choosing $a_1 = (2\varepsilon)^{-1}$ and $a_n = 0$ for all $n > 1$. 
Replacing~\eref{eq:rho_y_analyt} in Eqs.~\eref{eq:var_momexp} and~\eref{eq:epr_momexp},
we treat $\sigma_y^2$ and $\dot{S}_{\rm c}$ as functionals of $U$
and we define a suitable cost function
\begin{equation}\label{eq:loss_fn}
    \mathcal{L}[U] = \alpha \ell^2 \dot{S}_{\rm c}[U] + \frac{\sigma_y^2[U]}{\ell^2},
\end{equation}
again with $\ell = 2D_{\rm eff}/\kapt$, where the coefficient associated with $\sigma_y^2$ has been set to one without loss of generality. 
The $\ell$-dependent prefactors are designed to compensate for the different scalings of each term with $\ell$, cf.~Eqs.~\eref{eq:O_nabla_kappa_sigma} and~\eref{eq:O_nabla_kappa_epr}. This definition of the cost function formalizes the intuition that a desirable policy should simultaneously minimize the displacement from the target path and the excess entropy production associated with navigation. The relative importance of these two features is determined by the magnitude of the positive penalty factor $\alpha$.

\begin{figure}
    \centering
    \includegraphics[width=0.9\columnwidth]{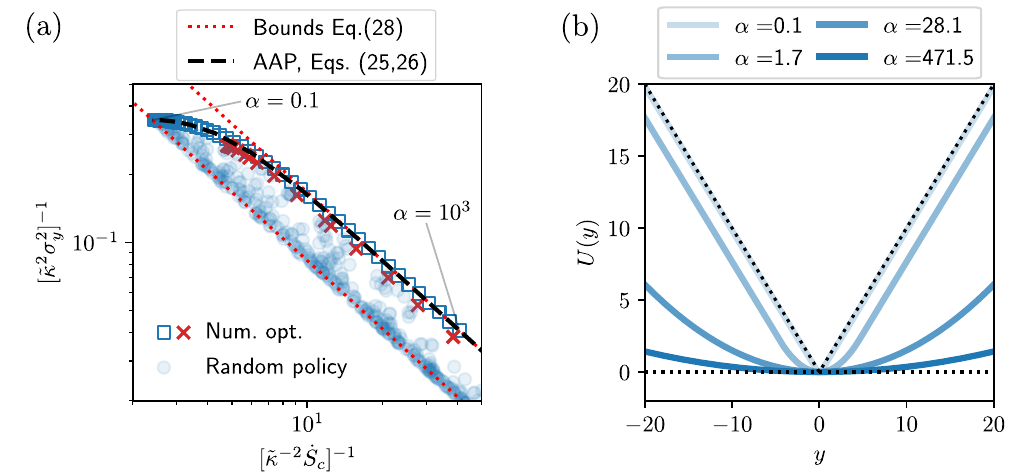}
    \caption{Numerical policy optimization and optimality of AAP. 
    (a) Performance comparison of random policies obtained by uniform sampling of the policy space $\{\log a_n\}_{1 \le n \le 3}$ (blue circles) vs optimal policies resulting from numerical minimization of the loss function defined in Eq.~\eref{eq:loss_fn} (blue squares) for $\tilde{\kappa}=0.1$ and ${\rm Pe} =10$. The numerical optima overlap with the performance curve of AAP for varying $\varepsilon$ (black dashed), as given by Eqs.~(\ref{eq:O_nabla_kappa_sigma},\ref{eq:O_nabla_kappa_epr}). Numerical optima obtained with the additional constrain $a_1=0$, shown as red crossed, visibly fall short of the Pareto front. Together, these results suggest that AAP is optimal. 
    (b) Example outcomes of numerical policy optimization based on the cost function Eq.~\eref{eq:loss_fn} for different values of $\alpha$. Dashed lines indicate the extreme cases $a_{1 \le n \le 3} \to \infty$ and $a_{1 \le n \le 3} \to 0$.}
    \label{fig:opt_prot}
\end{figure}

The optimal steering policy hence corresponds to the set of coefficients $\{a_n\}_{n \ge 1}$ that minimizes ${\cal L}$ for a given choice of $\alpha$. Since aligning policies focus trajectories around $y = 0$ where low order contributions in~\eref{eq:def_tildeU} dominate, 
we truncate the sum up to order six, i.e. we set $M = 3$.
We performed the optimization numerically using the Scipy implementation of the Nelder-Mead method~\cite{gao2012implementing},
and summarize the results in Fig.~\ref{fig:opt_prot}.
As expected, high values of $\alpha$ lead to an increase of the penalty for dissipation, resulting in shallower optimal protocols (Fig.~\ref{fig:opt_prot}(b)).
Plotting against each other the set of optimal $\sigma^2_y$ and $\dot{S}_{\rm c}$ obtained for various $\alpha$ in Fig.~\ref{fig:opt_prot}(a), 
we observe that the corresponding Pareto front remarkably coincides with the data obtained from AAP (Eqs.~(\ref{eq:O_nabla_kappa_sigma},\ref{eq:O_nabla_kappa_epr})) when varying $\varepsilon$.
In contrast, optimal protocols obtained under the constraint that $a_1 = 0$ fall short of the Pareto front. 
This indicates AAP's near-optimality in achieving localisation at minimum energy expenditure.
Exploring a set of random policies by sampling $\{\log a_n\}_{1 \le n \le 3}$ uniformly in $[-10,0]$, 
we further show in Fig.~\ref{fig:opt_prot}(a) that they all satisfy the inequalities derived in Eq.~\eref{eq:ineq_prod} for AAP, 
thus highlighting the generality of these bounds.

\subsection{Compensating aligning policies}\label{s:caap}

\begin{figure}
    \centering
    \includegraphics[width=0.9\textwidth]{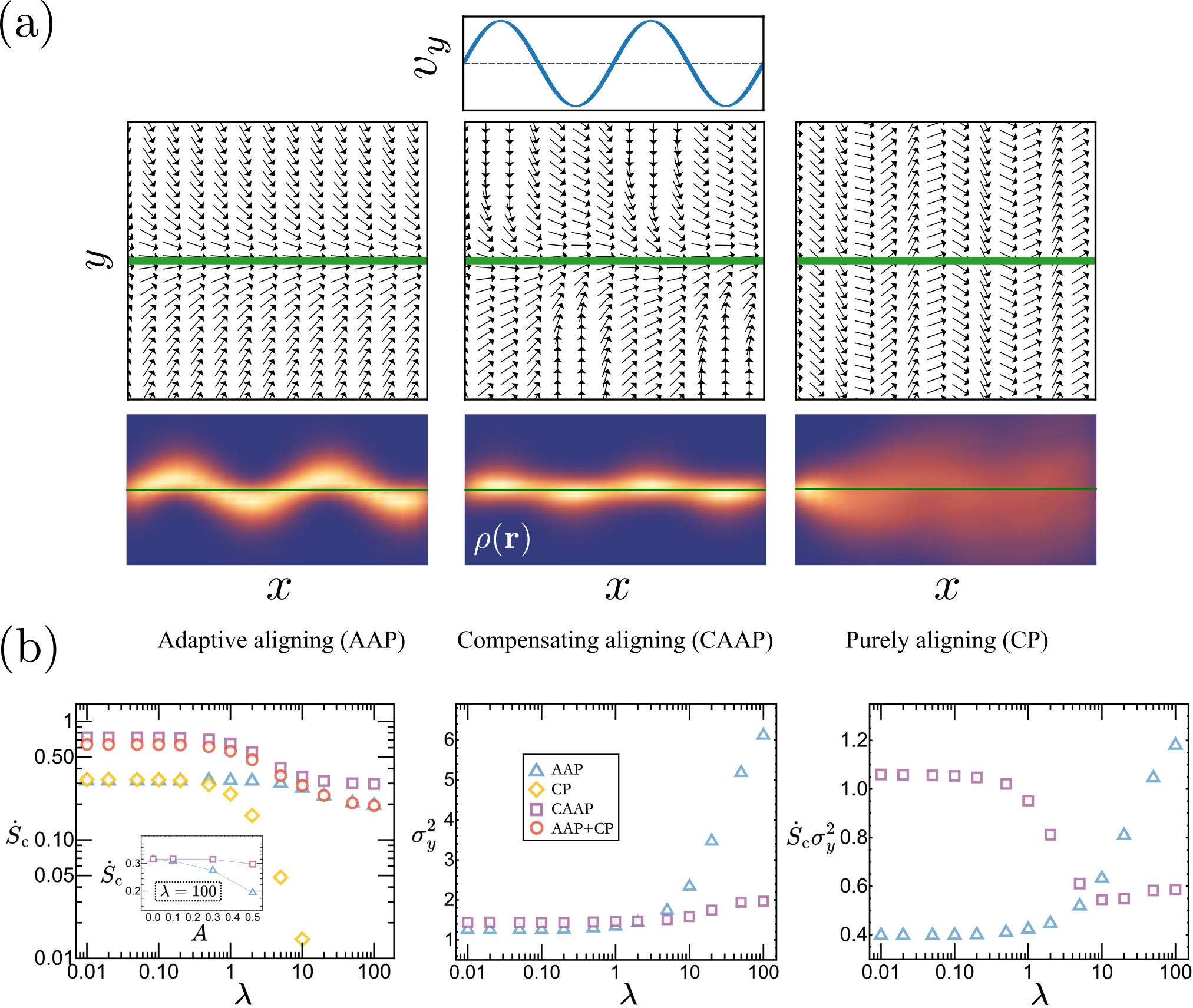}
    \caption{Localisation at a target path in the presence of a transverse sinusoidal flow for three examples of steering policies. (a) From left to right: adaptive aligning policy (AAP, cf.~Sec.~\ref{ss:aap}), compensating aligning policy (CAAP, cf.~Sec.~\ref{s:caap}), and purely compensating policy (CP, cf.~\ref{a:cp}). 
    The heatmaps show the steady-state empirical density normalised over the range $x \in [0,2\lambda]$ with $\lambda=20$. For the case of CP, where no steady-state exists due to uncompensated diffusion in the $y$ direction, we shift empirical trajectories such that $y(t_\lambda)=0$ where $t_\lambda: x(t_\lambda)=2n\lambda$ for $n\in \mathbb{Z}$ for visualisation.
    (b) Self-steering component of the mean entropy production rate, variance of the orthogonal displacement, and the product of the two as a function of the sinusoidal flow wavelength for the three policies listed in panel (a). The inset shows convergence of $\dot{S}_c$ for AAP and CAAP as the flow strength vanishes, $A \to 0$, as well as a weak dependence of the CAAP cost on $A$.
    In both panels, $\Pe=10$, $\varepsilon=1$, $\kapt=2$, and $A = 0.5$.
    }
    \label{fig:quiver_policies}
\end{figure}

So far, in this section, we have focused on localisation in the absence of external flows. We are now interested in how our previous findings are modified when navigation happens in the presence of a non-vanishing flow $\bm{v}(\bm{r})$ in Eq.~\eref{eq:lang_r}. 
Indeed, motion at low Reynolds number by real biological and artificial micro-machines typically takes place in complex environments \cite{bechinger2016active,tsang2020review} and the ability to reliably adapt to external flows might constitute an essential requirement for reaching a desired destination.
This naturally leads to the study of another class of policies $\hat{\bm{u}}^*(\bm{r},\bm{v}(\bm{r}))$ that depend also on the local flow velocity. Following \cite{piro2023optimal}, we refer to these as \emph{compensating aligning} policies.

For $\bm{v} \neq \bm{0}$, an optimal strategy might then intuitively be expected to leverage information about the flow landscape to improve localisation. In particular, in scenarios where the flow tends to advect the particle away from the target subspace, the associated policy should `compensate' for this effect, i.e.\ we expect $\sin\theta^*(\bm{r}) \approx -v_y(\bm{r})$ in units of the self propulsion velocity.
Variations in the flow that affect the policy performances are thus transverse to the path and vary along it. Henceforth, we choose for simplicity a sinusoidal flow profile characterized by a single length scale $\lambda$, $v_y(x) = A \sin(2 \pi x/\lambda)$.\\



Let us now specialise to the particular case where $\hat{\bm{u}}^*(y,\bm{v}(\bm{r})) = (\cos\theta^*,\sin\theta^*)$ satisfies

\begin{eqnarray}
    \sin\theta^*(y,\bm{v}(\bm{r})) &= \left\{\begin{array}{cc}
        1, &  y + \varepsilon v_y(\bm{r}) \le -\varepsilon \\
        -v_y(\bm{r})-y/\varepsilon, & |y + \varepsilon v_y(\bm{r})| < \varepsilon \\
        -1, & y + \varepsilon v_y(\bm{r}) \ge \varepsilon
    \end{array}\right.~, \nonumber \\
    \cos\theta^*(y,\bm{v}(\bm{r})) &= \sqrt{1-\sin^2\theta^*} ~, \label{eq:caap_def}
\end{eqnarray}
and where we have assumed a weak flow, $2A/\kapt<1$ in units of the self-propulsion velocity, to ensure localisation even in the weak alignment regime.
This corresponds to the compensating adaptive aligning policy (CAAP) introduced in Ref.~\cite{piro2023optimal}. Under ideal conditions of instantaneous reorientation, it is designed so as to exactly compensate for the effect of advection in the orthogonal direction while counteracting diffusion in the same spirit as AAP of Sec.~\ref{ss:aap}.

Comparing the AAP and CAAP performance in numerical simulations of the full Langevin dynamics under identical flow conditions, we observe that localisation is improved for sufficiently large $\lambda$ upon introduction of compensation (see Fig.~\ref{fig:quiver_policies}(a)). The comparatively poorer performance of CAAP at small $\lambda$ can be rationalised by noting that finite $\kapt$ introduces a delay in adaptation. Compensating for the flow variations then only makes sense when the swimmer can adapt faster than the typical variations in flow intensity and directions it experiences. 
Although CAAP manages to keep the displacement variance nearly independent of $\lambda$, the associated thermodynamic cost grows as $\lambda$ is reduced (Fig.~\ref{fig:quiver_policies}(b)). Again, this makes sense as small values of $\lambda$ will lead to an increased frequency of required reorientations. Note that the thermodynamic cost of CAAP is systematically larger than that of AAP, which both have a comparable accuracy at small $\lambda$.
In the spirit of Eqs.~\eref{eq:tradeoff_1pt_noflow} and~\eref{eq:ineq_prod}, we further illustrate the 
impact of compensation on the dissipation-accuracy tradeoff by plotting $\dot{S}_c \sigma_y^2$ against $\lambda$, Fig.~\ref{fig:quiver_policies}(c),  whereby we conclude that attempting to compensate for quickly varying environmental flows is actually counterproductive.

To break down the contributions to the dissipation arising in CAAP from compensation and alignment, 
we introduce a \emph{purely compensating} policy (CP), for which $\sin\theta^*(\bm{r}) = -v_y(\bm{r})$.
Although CP does not constrain the particle in the vicinity of the target path, it still causes dissipation as a result of the nonuniform flow landscape.
Like AAP, CP can be tackled analytically within the expansion presented in Sec.~\ref{ss:moment_exp_brief}, and we show in \ref{a:cp} that $\dot{S}_{\rm c}^{\rm CP} = {\cal O}(1)$ for $\lambda \to 0$, while $\dot{S}_{\rm c}^{\rm CP} = \lambda^{-2}$ when $\lambda \to \infty$.
Comparing the dissipation associated with the three policies, the data presented in Fig.~\ref{fig:quiver_policies}(b) further suggests that $\dot{S}_{\rm c}^{\rm CAAP}$ is bounded from below by $\dot{S}_{\rm c}^{\rm AAP} + \dot{S}_{\rm c}^{\rm CP}$ for all values of $\lambda$. 
While for $\lambda \to 0$ this bound appears tight, for slowly varying flows $\dot{S}_{\rm c}^{\rm CP} + \dot{S}_{\rm c}^{\rm AAP} \simeq \dot{S}_{\rm c}^{\rm AAP}$ is appreciably lower than $\dot{S}_{\rm c}^{\rm CAAP}$, as a result of the fact that CAAP achieves significantly better precision than AAP in this regime. 

\begin{figure}
    \centering
    \includegraphics[width=0.7\columnwidth]{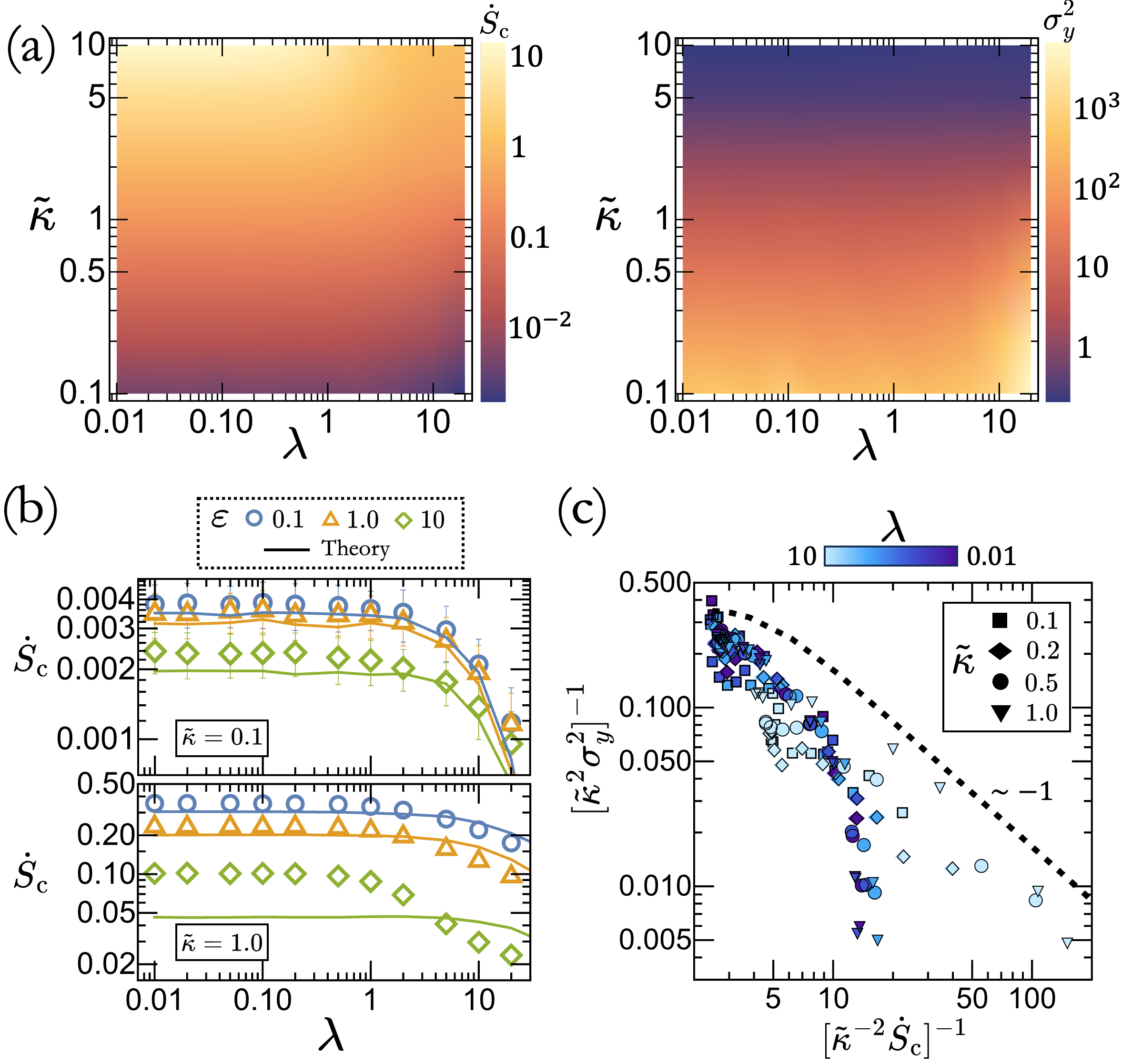}
    \caption{Numerical simulations of CAAP in the presence of a sinusoidal flow. (a) Color maps of the entropy production rate (left panel) and transversal variance (right panel) as a function of the alignment strength $\kapt$ and of flow wavelength $\lambda$. The confinement scale is here set to $\varepsilon=1$. (b) Comparison between the values of the entropy production rate obtained from the numerical simulations (symbols) and the ones predicted by Eq.~\eref{eq:sdot_c_11mom} using empirical density profiles (solid lines) as a function of $\lambda$. Different colors here indicate the value of $\varepsilon$ as per the legend while each panel corresponds to a value of $\kapt$. All data shown in (a) and (b) have been obtained at $\Pe=10$ and for a flow amplitude $A=0.5$ in units of the swimmer's self-propulsion speed. (c) Rescaled entropy production and variance for different values of $\kapt$, $\varepsilon$ and $\lambda$. The dashed line indicates the AAP/CAAP performance line in the absence of flow, Eqs.~(\ref{eq:O_nabla_kappa_sigma},\ref{eq:O_nabla_kappa_epr}).}
    \label{fig:CAAP}
\end{figure}

We now proceed to a more systematic exploration of CAAP for the sinusoidal flow introduced above, the results of which are summarized in Fig.~\ref{fig:CAAP}.
We show in Fig.~\ref{fig:CAAP}(a) the mean rate of entropy production and variance of the orthogonal displacement from the target path evaluated from Langevin simulations as a function of $\lambda$ and $\kapt$. Clearly, a dissipation-accuracy tradeoff is at play here, with $\dot{S}_{\rm c}$ increasing in regions of low $\sigma_y^2$ and vice versa. 
As we could not obtain an analytical expression for the full steady state density, we furthermore compare our numerical estimates of $\dot{S}_{\rm c}$ with 
the prediction~\eref{eq:sdot_c_11mom} using empirical density profiles to perform spatial averaging.
As shown in Fig.~\ref{fig:CAAP}(b), Eq.~\eref{eq:sdot_c_11mom} works well at small $\kapt$ and large $\lambda$, where our expansion applies. 
Both the predicted and measured $\dot{S}_{\rm c}$ are found to increase monotonically as $\lambda$ decreases, consistent with the intuition that adaptation to a faster-varying environment should incur a higher thermodynamic cost. 
A scatter plot of the numerically estimated $\dot{S}_{\rm c}$ and $\sigma_y^2$ for different values of $\kapt \lesssim 1$, $\varepsilon$ and $\lambda$ is shown in Fig.~\ref{fig:CAAP}(c).
In particular, we compare these points with the relation derived for AAP/CAAP in the absence of flow, Eqs.~(\ref{eq:O_nabla_kappa_sigma},\ref{eq:O_nabla_kappa_epr}) (note that the two policies are equivalent when $A = 0$, as illustrated in Fig.~\ref{fig:quiver_policies}(b)).
Clearly, the requirement to compensate for transverse flow components generally worsens the dissipation-accuracy tradeoff, while this effect intensifies with decreasing $\lambda$ in line with the above observations.

\section{Discussion and conclusion}\label{s:discussion_conclusion}

Establishing a framework to characterise the out-of-equilibrium behaviour of systems integrating force generation with environmental sensing and information processing constitutes a crucial open challenge for the field of active matter~\cite{levine2023physics,goldman2024robot,BowickPRX2022}. In particular, minimal models shall play a key role in this endeavour by enabling the identification fundamental constraints and tradeoffs that underly the operation of living matter. 

In this work, we have considered a minimal model of an active Brownian particle able to actively control its self-propulsion orientation according to a steering policy designed to localise it in a specific region of space. By quantifying the thermodynamic cost of active reorientations via a suitable decomposition of the average rate of entropy production~\cite{seifert2012stochastic,cocconi2020entropy}, we uncovered a generic dissipation-accuracy tradeoff at play in this context. We have illustrated this tradeoff in several scenarios, including localisation at a point target (Section~\ref{s:pt_target}), and navigation along a target path (Section~\ref{s:target_path}), contrasting performance in the presence and absence of external flows.

In the absence of flows and within the weak alignment regime, we computed in Section~\ref{sec_optimal_policy} the set of non-dominated solutions for which accuracy of the localisation can only be improved at the expense additional dissipation, known as Pareto front~\cite{ngatchou2005pareto}. Interestingly, the measures of accuracy and dissipation for this set of optimal policies coincide with those of the adaptive aligning policy (AAP) (see Fig.~\ref{fig:opt_prot}(a)), which was introduced in~\cite{piro2022optimal} as an approximate travel time minimising navigation policy that allows for autonomous control. In future works, it will be interesting to explore further the robustness of policies like AAP when subjected to additional constraints on viscous dissipation or limited sensing abilities, as well as beyond the weak alignment regime.

The treatment proposed here does not take into account rotations imposed on the self-propulsion orientation originating by gradients of the external fluid flow ---which depend on the swimmer's body shape~\cite{piro2024energetic}--- as well as the possible presence of potential forces that are typically at play in complex environments~\cite{piro2022efficiency,VolpePNAS2017}, e.g.\ due to obstacles or inhomogeneous substrates. The inclusion of these effects is likely to affect the dissipation in nontrivial ways and qualitatively affect its relation to accuracy. 

Other open questions concern the consequences of allowing the active particle to control not only the orientation but also the magnitude of its fluctuating self-propulsion force~\cite{piro2023optimal,auschra2021_nudging,mahault2023emergent}, 
and how this situation differs from the conceptually related scenario where the control is external~\cite{davis2023active.14.011012,Proesmans2023}.
Taking into account additional thermodynamic costs associated with sensing, the acquisition of a policy, e.g. via autonomous learning \cite{stark2021artificial,schneider2019optimal,muinos2021reinforcement,biferale2019zermelo,putzke2023optimal}, or its negotiation across agents, e.g. in the context of shape assembly~\cite{sun2023mean,rubenstein2014programmable,slavkov2018morphogenesis}, will enable extensions to the context of multi-agent interactions, both competitive and cooperative. 
Finally, important insight is likely to be gained by developing thermodynamically consistent coarse-grained models of smart active particles in the spirit of recent works such as Refs.~\cite{chatzittofi2023entropy,fritz2023thermodynamically,speck2019thermodynamic,bebon2024thermodynamics}, where effective active forces and torques are done away with in favour of microscopically resolved dissipative currents of auxiliary reactants and products.\\

\textbf{Acknowledgements:} L.\ C.\ thanks Letian Chen for useful discussion on an earlier iteration of the model presented in Appendix B. We acknowledge support from the Alexander von Humboldt Foundation. This work was supported by the European Research Council (ERC) under the European Union’s Horizon 2020 research and innovation programme (Grant Agreement No. 882340).

\appendix

\section{The coarse-graining approach}\label{a:mom_exp_detail}

In this appendix, we present the derivation of the approximation scheme introduced in Sec.~\ref{ss:moment_exp_brief} to determine the steady-state solution of the dimensionless Fokker-Planck equation \eref{eq:fp_full}. Starting from an assumed adiabatic limit for the angular dynamics and drawing on a systematic gradient expansion, this procedure allows us to iteratively obtain the angular moments of the joint distribution $P(\bm{r},\theta,t)$.

\subsection{The general scheme}

Starting from the dimensionless Fokker-Planck equation
\begin{equation} \label{eq_FP}
\partial_t P + \nabla\cdot \left[ \left( \hu(\theta) + \bm v - \Pe^{-1}\nabla \right) P \right] + \partial_\theta \left[  \left( \kapt \Gamma(\theta,\bm r) - \partial_\theta \right) P \right] = 0,
\end{equation}
we seek, in the spirit of an adiabatic approximation, factorised solutions of the form~\cite{mahault2023emergent}
\begin{equation*} 
P(\bm r,\theta,t) = \rho(\bm r,t) Q(\theta,\bm r) = \rho(\bm r,t) \left[ Q_0(\theta,\bm r) + Q_1(\theta,\bm r) + \ldots \right],
\end{equation*}
where $Q_i = {\cal O}(\nabla^i)$ are of increasing order in the spatial gradients.
The dynamics of the density field $\rho$ is obtained after integrating Eq.~\eref{eq_FP} over the orientation $\theta$, namely
\begin{equation} \label{eq_rhoeq}
\partial_t \rho + \nabla\cdot \left[ \left( \langle \hu \rangle_{\theta} + \bm v - \Pe^{-1}\nabla \right) \rho \right] = 0,
\end{equation}
where we have defined 
\begin{equation*}
\langle \cdot \rangle_{\theta} \equiv \rho^{-1}\int_0^{2\pi}\rmd\theta \, (\cdot) P = \int_0^{2\pi}\rmd\theta \, (\cdot) Q~.
\end{equation*}
For later use, we also introduce the notation $\langle \cdot \rangle_{k} \equiv \int_0^{2\pi}\rmd\theta \, (\cdot) Q_k$ for averages computed using $Q_k$ as a measure. 
Combining Eqs.~\eref{eq_FP} and \eref{eq_rhoeq}, we obtain
\begin{eqnarray} 
Q\partial_t \rho 
&= - Q \nabla\cdot \left[ \left( \langle \hu \rangle_{\theta} + \bm v - \Pe^{-1}\nabla \right) \rho \right]  \nonumber \\
&= -\nabla\cdot \left[ \left( \hu(\theta) + \bm v - \Pe^{-1}\nabla \right) Q \rho \right] - \rho \partial_\theta \left[  \left( \kapt\Gamma(\theta,\bm r) - \partial_\theta \right) Q \right].\label{eq_rhoQ}
\end{eqnarray}
Expressed at zeroth order in gradients, Eq.~\eref{eq_rhoQ} reduces to
\begin{equation*}
\rho\partial_\theta \left[  \left( \kapt \Gamma(\theta,\bm r) - \partial_\theta \right) Q_0 \right] = 0,
\end{equation*}
whence
\begin{equation} \label{eq_Q_0}
Q_0(\theta,\bm r) = \frac{e^{\kapt U_\theta(\theta,\bm r)}}{Z_0(\bm r)}~,
\end{equation}
where the generalized potential satisfies $\partial_\theta U_\theta(\theta,\bm r) = \Gamma(\theta,\bm r)$ and $Z_0(\bm r) \equiv \int\rmd\theta \exp[\kapt U_\theta(\theta,\bm r)]$ is a normalization constant.
Expanding $Q$ in~\eref{eq_rhoQ} and equating terms of same order in gradients, we obtain, for $n > 0$,
\begin{eqnarray} \label{eq_rhoQn}
\fl \rho \partial_\theta \left[  \left( \kapt\Gamma(\theta,\bm r) - \partial_\theta \right) Q_n \right]
& = \sum_{k=0}^{n-1} Q_{n-k-1}\nabla\cdot \left[ \langle \hu \rangle_k \rho \right] + Q_{n-1} \nabla\cdot \left[ \bm v \rho \right] - \nabla\cdot \left[ \left( \hu(\theta) + \bm v \right) Q_{n-1} \rho \right] \nonumber \\
& + \Pe^{-1}\left[ \Delta(\rho Q_{n-2}) - Q_{n-2}\Delta\rho \right].
\end{eqnarray}
To simplify the left-hand side of~\eref{eq_rhoQn}, we write $Q_n \equiv Q_0 \calX_n$, which leads to 
\begin{equation*}
\partial_\theta \left[  \left( \kapt\Gamma(\theta,\bm r) - \partial_\theta \right) Q_n \right] = -\partial_\theta(Q_0 \partial_\theta\calX_n).
\end{equation*}
The right-hand side, in turn, can be recast as
\begin{eqnarray}
 \fl \rho\calB_n(\theta,\bm r) & \equiv -\sum_{k=0}^{n-1} Q_{n-k-1}\nabla\cdot \left[ \langle \hu \rangle_k \rho \right] 
 - Q_{n-1} \nabla\cdot \left[ \bm v \rho \right] 
 + \nabla\cdot \left[ \left( \hu(\theta) + \bm v \right) Q_{n-1} \rho \right] \nonumber \\
\fl  & - \Pe^{-1}\left[ \Delta(\rho Q_{n-2}) - Q_{n-2}\Delta\rho \right]  \nonumber \\
\fl & = \nabla\cdot\left[ \left( \hu(\theta) - \langle \hu \rangle_{0} \right) \rho Q_{n-1} \right] 
+ \rho \left( \langle \hu \rangle_{0} + \bm v \right)\cdot \nabla Q_{n-1}
- \sum_{k=1}^{n-1} Q_{n-k-1}\nabla\cdot \left[ \langle \hu \rangle_k \rho \right]  \nonumber \\
\fl & - \Pe^{-1}\left[ \rho\Delta Q_{n-2} + 2(\nabla Q_{n-2})\cdot (\nabla\rho) \right]~,
\label{eq_Bn}
\end{eqnarray}
which only depends on $Q_m$, and thus $\calX_m$, with $m < n$.
Therefore, we rewrite~\eref{eq_rhoQn} compactly as
\begin{equation}\label{eq_rhoQ_On_simpler}
 \partial_\theta(Q_0 \partial_\theta\calX_n) = \calB_n (\theta,\bm r)~,
\end{equation}
which constitutes the basis of our iterative scheme.
The solution of Eq.~\eref{eq_rhoQ_On_simpler} for $\calX_n$ if found by straightforward integration to take the form
\begin{equation} \label{eq_solXn}
\calX_n(\theta,\bm r) = \calY_n(\bm r) + \int_0^{\theta}\frac{\rmd\theta'}{Q_0(\theta',\bm r)} \left[ \calZ_n(\bm r) + \int_0^{\theta'}\rmd\theta'' \, \calB_n(\theta'',\bm r) \right].
\end{equation}
To fix the two constants of integration, we first note that $\calX_n$ must be $2\pi$-periodic in the orientation $\theta$, which imposes that
 \begin{equation*}
 \calZ_n(\bm r) = - \frac{\int_0^{2\pi}\frac{\rmd \theta}{ Q_0(\theta,\bm r)} \int_0^{\theta}\rmd\theta' \, \calB_n(\theta',\bm r)}{\int_0^{2\pi}\frac{\rmd \theta}{Q_0(\theta,\bm r)}}.
 \end{equation*}
 The second condition, $\int_0^{2\pi}\rmd\theta\, Q_n = \langle \calX_n \rangle_0 = 0$ $\forall n > 0$, results from the fact that the marginal distribution $Q$ must be normalized.
 Hence, we have
 \begin{equation*}
 \calY_n(\bm r) = - \left\langle \int_0^{\theta}\frac{\rmd\theta'}{Q_0(\theta',\bm r)} \left[ \calZ_n(\bm r) + \int_0^{\theta'}\rmd\theta'' \, \calB_n(\theta'',\bm r) \right] \right\rangle_0.
 \end{equation*}
 Crucially, the solution~\eref{eq_solXn} allows one to obtain the expression of all orientational moments of $P$ at any order in $\nabla$.
 Namely, for any function $\calM(\theta,\bm r)$, 
\begin{equation}
\langle \calM(\theta,\bm r)\rangle_\theta = \langle \calM\rangle_0 + \sum_{n=1}^{\infty} \langle \calX_n \calM\rangle_0 .
\end{equation}
In practice, however, calculations become quickly complicated when considering high order gradient terms, forcing us to truncate the sum at finite order.

\subsection{Solution at ${\cal O}(\nabla)$}

\subsubsection{Rate of entropy production}
We first derive a general expression for the rate of entropy production associated with the feedback control in steady state, $\dot{S}_{\rm c}$, at leading order in the gradient expansion.
Starting from its definition~\eref{eq:epr_exact}, we write
\begin{eqnarray*}
\dot{S}_{\rm c} & = \int\rmd\theta\rmd\bm r\, \kapt\Gamma(\theta,\bm r) \rho_{\rm s}(\bm r) \left( \kapt \Gamma(\theta,\bm r) - \partial_\theta \right)Q_1(\theta,\bm r) + {\cal O}(\nabla^2)~, \nonumber \\
& = - \int\rmd\theta\rmd\bm r\, \kapt U_\theta(\theta,\bm r) \rho_{\rm s}(\bm r) \partial_\theta\left[ \left( \kapt \Gamma(\theta,\bm r) - \partial_\theta \right)Q_1(\theta,\bm r) \right] + {\cal O}(\nabla^2)~,
\end{eqnarray*}
where the contribution from $Q_0$ in the first line was discarded noting that the latter corresponds to the adiabatic approximation $J_\theta = 0$, 
while the second line was obtained after integrating by parts over $\theta$.
Using Eqs.~\eref{eq_Q_0} and~\eref{eq_rhoQ_On_simpler}, as well as the fact that $\int\rmd\theta\, \calB_1(\theta,\bm r) = 0$, we then obtain
\begin{equation} \label{eq_Sc_B1_O1}
\dot{S}_{\rm c} = \int\rmd\theta\rmd\bm r \, \ln[Q_0(\theta,\bm r)] \rho_{\rm s} \calB_1(\theta,\bm r) + {\cal O}(\nabla^2)~,
\end{equation} 
where $\calB_1(\theta,\bm r)$ is given by
\begin{equation*}
\rho_{\rm s} \calB_1(\theta,\bm r)
 = \nabla\cdot\left[ \left( \hu(\theta) - \langle \hu \rangle_{0} \right) \rho_{\rm s} Q_0 \right] 
 + \rho_{\rm s} \left( \langle \hu \rangle_{0} + \bm v \right)\cdot \nabla Q_0 .
\end{equation*}
At leading order in gradients, the integrand in the definition of $\dot{S}_{\rm c}$ is thus fully determined by the distribution $Q_0$.
After replacing $\calB_1$ by its expression and operating a few manipulations, we obtain 
\begin{eqnarray} 
\dot{S}_{\rm c} & = \kapt\int\rmd\bm r \rho_{\rm s}\left[ 
\kapt\left(\langle\hu\rangle_0 + \bm v\right)\cdot\left( \langle U_\theta\nabla U_\theta\rangle_0 - \langle U_\theta\rangle_0\langle \nabla U_\theta\rangle_0 \right)\right. \nonumber \\
& \left. - \left\langle \left(\hu - \langle \hu \rangle_0\right)\cdot\nabla U_\theta \right\rangle_0
\right] + {\cal O}(\nabla^2)~.
\label{eq_Sc_O1_general}
\end{eqnarray}

In the special case where the steering policy satisfies $\Gamma(\theta,\bm r) = \Gamma(\theta - \theta^*(\bm r))$,
we have $\nabla U_\theta = -\partial_\theta U_\theta \nabla\theta^*$, while for all $n \ge 0$,
\begin{equation*}
\langle U_\theta^n \partial_\theta U_\theta\rangle_0 = 
\frac{(-1)^{n+1} n!}{\kapt^{n+1} Z_0} \int \rmd \theta \, \partial_\theta \left(\exp[\kapt U_\theta]\right) = 0~,
\end{equation*}
which can be obtained via multiple integrations by parts.
Hence, when the steering protocol depends on space only through $\theta^*$, most of the contributions to $\dot{S}_{\rm c}$ vanish.
Similarly, we simplify the only remaining term on the right hand side of Eq.~\eref{eq_Sc_O1_general} by noting that
$\nabla \cdot \langle \hu \rangle_0 = \kapt \langle \hu \nabla U_\theta \rangle_0$.
Therefore, we finally get
\begin{equation} \label{eq_Sc_O1_final}
\dot{S}_{\rm c} = -\left\langle \nabla\cdot \langle \hu \rangle_0 \right\rangle + {\cal O}(\nabla^2)~,
\end{equation}
where angular brackets deprived of indices indicate that the average is taken with respect to the steady state density distribution $\rho_{\rm s}$.

\subsubsection{Steady state density field at ${\cal O}(\nabla,\kapt)$}

To calculate $\dot{S}_{\rm c}$ from~\eref{eq_Sc_O1_final}, we therefore have to determine the steady state density distribution solution of Eq.~\eref{eq_rhoeq}.
This equation, in particular, involves the average particle orientation with respect to the distribution $Q_1$, whose expression is formally given by Eq.~\eref{eq_solXn}.
In practice, however, evaluating explicitly $Q_1$ and the corresponding moments in the general case quickly leads to lengthy expressions.
Here, we therefore focus on the aligning function $\Gamma(\theta) = -\sin \theta$ used in the main text.
For this case, we have $U_\theta(\theta) = \cos\theta$, while $\langle \hu \rangle_0 = c_0(\kapt)\hu^*$ with $\hu^* = \hu(\theta^*)$ 
and  $c_0(\kapt) = |\langle \hu \rangle_0| = I_1(\kapt)/I_0(\kapt)$ where $I_n$ is the modified Bessel function of the first kind of order $n$.
The feedback control entropy production rate then reduces to 
\begin{equation} \label{eq_dotSc_app_mom_exp}
\dot{S}_{\rm c} = -c_0(\kapt)\left\langle \nabla\cdot \hu^* \right\rangle + {\cal O}(\nabla^2)~.
\end{equation}
In order to get analytically tractable expressions, below we further consider the limiting case $\kapt \ll 1$ of weak alignment, 
truncating the resulting expressions to leading order. 
As $c_0(\kapt) = \frac{\kapt}{2} + {\cal O}(\kapt^2)$, we recover the expression of the entropy production rate presented 
in Eq.~\eref{eq:sdot_c_11mom} of the main text.

For the full probability distribution, we find after some algebra
\begin{eqnarray}
\calX_1(\theta,\bm r) = &-\left[ \hu(\theta) - \frac{\kapt}{2}\hu(\theta^*) - \frac{\kapt}{8}\hu(2\theta-\theta^*) \right]\cdot\nabla\ln\rho \nonumber \\
&- \kapt \left[ \sin(\theta - \theta^*)\bm v - \frac{1}{8}\hu^{\perp}(2\theta - \theta^*) \right]\cdot\nabla\theta^*
+ {\cal O}(\kapt^2)~,
\label{eq_X1_app_mom_exp}
\end{eqnarray}
whereby
\begin{equation*} \label{eq_moment_e_O11}
\langle \hu \rangle_{\theta} = \frac{1}{2} \left[ \kapt (1 - \bm v\cdot\nabla)\hu(\theta^*) - \nabla \ln\rho  \right] + {\cal O}(\nabla^2,\kapt^2)~.
\end{equation*}
Replacing the above expression for the mean particle orientation in Eq.~\eref{eq:rho_dyn} of the main text, 
we finally recover the effective equation~\eref{eq:cg_fp_rho_main} for the dynamics of the particle density.

\subsection{${\cal O}(\nabla^2,\kapt)$ solution}
\label{app_Onabla2_exp}

In~\ref{a:cp} we study a compensating policy (CP) and show that for a rapidly varying flow the behavior of the entropy production is not faithfully captured by the leading order contribution~\eref{eq_dotSc_app_mom_exp}.
In this section, we therefore pursue the gradient expansion to the next order, while working at linear order in $\kapt$ for convenience.
As detailed above, we use the expression of $\calX_1$ determined in~\eref{eq_X1_app_mom_exp} to calculate $\calX_2$ via Eqs.~\eref{eq_Bn} and~\eref{eq_solXn}. 
Even at linear order in $\kapt$, the resulting formula for $\calX_2$ remains long and uninformative. 
Hence, we directly give write the ensuing contributions to the relevant moments.
Namely, the ${\cal O}(\nabla^2)$ correction to the self-propulsion orientation moment, $\langle \hu \rangle_2$, is given by 
\begin{eqnarray}
\fl \langle \hu \rangle_2 = &\frac{1}{2}(\bm v \cdot \nabla) \nabla\ln\rho \nonumber \\
\fl & + \frac{\kapt}{2}\left\{ 
-\hu^*\left[ \frac{1}{4}\left(\nabla^2\ln\rho + |\nabla\ln\rho|^2\right) + \left( \Pe^{-1} + \frac{1}{16} \right)|\nabla\theta^*|^2 + (\bm v \cdot \nabla\theta^*)^2\right] \right. \nonumber\\
\fl & \left. + \hu_\perp^* \left[ \left( \Pe^{-1} + \frac{1}{16} \right)\nabla^2\theta^* + (\bm v\cdot\nabla)[(\bm v\cdot \nabla)\theta^*] + 2\Pe^{-1}(\nabla\theta^*)\cdot(\nabla\ln\rho) \right] \right.  \nonumber \\
\fl & \left. + \frac{1}{8} \bigg[ 2 [(\nabla\ln\rho) \cdot \nabla]\hu^* - (\hu^* \cdot \nabla)\nabla\ln\rho - 5 [(\hu^* \cdot \nabla)\ln\rho]\nabla\ln\rho \right. \nonumber \\
\fl & \left. - 3 [\nabla\cdot \hu^*]\nabla\ln\rho - 2[\nabla \hat{u}_i^*]\partial_i\ln\rho \bigg] 
\right\} ~,
\label{eq_moment_e_O21}
\end{eqnarray}
where sum over repeated roman indices is implied.
In turn, the moments involved in Eq.~\eref{eq:dots_c} read
\begin{eqnarray}
\fl & \langle \cos(\theta-\theta^*) \rangle_2 = \frac{1}{2}\left( \hu^*\bm v : \nabla\nabla \right)\ln\rho
- \frac{\kapt}{2}\left[ \left( \Pe^{-1} + \frac{1}{16} \right)|\nabla\theta^*|^2 + \frac{1}{2}(\hu^* \cdot \nabla \ln\rho)^2 \right.  \\
\fl & \left. + \frac{5}{16\rho}\nabla^2\rho + (\bm v \cdot \nabla\theta^*)^2 + \frac{1}{8\rho}(\bm q^*:\nabla\nabla)\rho + \frac{1}{8}(\nabla\ln\rho\times\nabla\theta^*)\cdot\hat{\bm z} 
+ \frac{1}{4} [\nabla\cdot \bm q^*]\cdot\nabla\ln\rho\right], \nonumber \\
\fl & \langle \sin^2(\theta-\theta^*) \rangle_2 = -\frac{1}{16\rho}(\bm q^*:\nabla\nabla)\rho,
\end{eqnarray}
where $\bm q^* = \hu^*\hu^* - \bm I/2$ denotes the symmetric traceless nematic tensor built from $\hu^*$, while $\bm A:\bm B\equiv {\rm Tr}(\bm A\bm B)$ for two matrices $\bm A$ and $\bm B$.
Gathering these results, computing the average with respect to the steady state density and discarding boundary terms, we finally obtain for the entropy production rate
\begin{eqnarray}
 \dot{S}_{\rm c} & = \frac{\kapt}{2}\bigg[ -\langle \nabla \cdot \hu^*\rangle - \left\langle \hu^*\bm v : \nabla\nabla \ln\rho\right\rangle
+ \kapt\left( \Pe^{-1} + \frac{1}{16} \right)\left\langle|\nabla\theta^*|^2\right\rangle  \nonumber \\
& + \kapt\left\langle(\bm v \cdot \nabla\theta^*)^2\right\rangle - \frac{\kapt}{4} \left\langle \nabla\nabla : \bm q^* \right\rangle - \frac{\kapt}{2} \left\langle \nabla\cdot[\hu^* (\hu^*\cdot\nabla) \ln\rho]\right\rangle \bigg]. \label{eq:epr_corr_k1_n2_app}
\end{eqnarray}

For the particular case of the purely compensating policy studied in Sec.~\ref{a:cp} , we have $\bm v(\bm r) = (v_x=0,v_y(x))$, whence $\partial_x v_x = 0$, $\partial_y v_y =0$ and $\partial_y \theta^*(v_y(x)) = 0$. Further, by translational symmetry along the coordinate $y$ orthogonal to the target path, the existence of a steady-state density can be enforced by taking periodic boundary conditions for $y$ in a narrow strip around the path, whereby at steady state $\partial_y \rho=0$ as well.
Then, some of the terms in \eref{eq:epr_corr_k1_n2_app} vanish, namely
\begin{eqnarray*}
    \fl \left\langle \hu^*\bm v : \nabla\nabla \ln\rho\right\rangle = \langle \cos\theta^* v_x \partial_x^2 \ln \rho \rangle = 0~, \quad {\rm and} \quad
    \left\langle(\bm v \cdot \nabla\theta^*)^2\right\rangle = \langle (v_x \partial_x \theta^*)^2 \rangle = 0~.
\end{eqnarray*}
For the others, we obtain the simplified forms
\begin{eqnarray*}
\fl  & \left\langle|\nabla\theta^*|^2\right\rangle = \left\langle|\partial_x\theta^*|^2\right\rangle , \\
\fl  &  - \left\langle \nabla\nabla : \bm q^* \right\rangle = - \left\langle \partial_x^2 \left[\cos^2\theta^* - \frac{1}{2} \right] \right\rangle  
 = \langle 2 \cos(2\theta^*)|\partial_x \theta^*|^2 + \sin(2\theta^*)(\partial_x^2 \theta^*)\rangle  \\
\fl  &  - \left\langle \nabla\cdot[\hu^* (\hu^* \cdot\nabla) \ln\rho]\right\rangle =  - \left\langle \partial_x [\cos^2\theta^* \partial_x \ln \rho] \right\rangle 
    = \langle \partial_x^2 (\cos^2\theta^*)\rangle -  \langle \cos^2\theta^* \partial_x^2 \ln \rho \rangle
    \nonumber \\
\fl  & \qquad\qquad\qquad\qquad\qquad\;\;\, = \left[ \left\langle 2\cos(2\theta^*)|\partial_x \theta^*|^2 +  \sin(2\theta^*)(\partial_x^2 \theta^*) \right\rangle - \langle \cos^2\theta^* \partial_x^2 \ln \rho \rangle \right]~.
\end{eqnarray*}
Hence, for this particular scenario,
\begin{eqnarray}
 \dot{S}_{\rm c} & = -\frac{\kapt}{2}\left[ \langle\partial_x \cos\theta^* \rangle 
 - \kapt\left( \Pe^{-1} + \frac{1}{16} \right) \left\langle|\partial_x\theta^*|^2\right\rangle \nonumber \right. \\
 &
 \left. + \frac{\kapt}{2} \left\langle 
 \cos^2\theta^* \partial_x^2 (\ln \rho) 
 - 3 \cos(2\theta^*)|\partial_x \theta^*|^2 - \frac{1}{2}\sin(2\theta^*)(\partial_x^2 \theta^*)
 \right\rangle \right]. \label{eq:epr_corr_k1_n2_app_CP}
\end{eqnarray}
As for Eq.~\eref{eq_dotSc_app_mom_exp},
Eq.~\eref{eq:epr_corr_k1_n2_app_CP} allows for the evaluation of $\dot{S}_{\rm c}$ from direct measurements of the steady state particle density.
In Fig.~\ref{fig:CP_plots}, we use Eq~\eref{eq:epr_corr_k1_n2_app_CP}, together with the steady state density calculated in Eq.~\eref{eq:sol_dens_driv}, to theoretically predict the entropy production.

\section{A point target in a uniform flow} \label{a:target_uniform}

In this appendix, we address the case of localisation at a point target located at $r=0$ in the presence of a uniform longitudinal flow, $\bm{v} = v_{\rm f} \hat{\bm{x}}$ with policy $\hu^* = -\hat{\bm r}$. 
Here, we use polar coordinates $(x,y) \to (r,\phi)$ and denote the polar basis vectors as $\hat{\bm{r}}$ and $\hat{\bm{\phi}}$, whence $\hat{\bm{x}} = \hat{\bm{r}}\cos\phi - \hat{\bm{\phi}}\sin\phi$.
This scenario is a relevant approximation in cases where a swimmer is required to localise within a typical radius from the target which is small compared to the characteristic length scale of the flow.
To solve Eq.~\eref{eq:cg_fp_rho_main} for the steady state swimmer density, we rewrite it as
\begin{equation}
   D_{\rm eff} \left( \hat{\bm{r}} \partial_r + \frac{\hat{\bm{\phi}}}{r} \partial_\phi \right) \ln \rho_{\rm s} = v_{\rm f} (\hat{\bm{r}}\cos\phi - \hat{\bm{\phi}}\sin\phi) - \frac{\kapt}{2} 
    \left(
    \hat{\bm{r}}  + \frac{v_{\rm f} \sin\phi}{r} \hat{\bm{\phi}}
    \right)~.
\end{equation}
Solving the two resulting equations, we eventually find
\begin{equation} \label{eq_density_uniform}
    \rho_{\rm s}(r,\phi) = \mathcal{N} {\rm exp}\left[ - \frac{r}{\ell} + \frac{v_{\rm f}\cos\phi}{\ell} \left( \frac{2r}{\kapt} + 1 \right) \right]
\end{equation}
with $\mathcal{N}$ a normalisation factor and $\ell = 2D_{\rm eff}/\kapt$ as defined in Sec.~\ref{s:pt_target}. 
Evaluating~\eref{eq_density_uniform} along the $y=0$ line by setting $\phi = \pi$, for $x<0$, and $\phi=0$, for $x>0$, we find that the density exhibits exponential tails with characteristic length scales: $\ell_{x<0} = \ell(1+2v_{\rm f}/\kapt)^{-1}$ and $\ell_{x>0} = \ell(1-2v_{\rm f}/\kapt)^{-1}$. 
Thus, similarly to the case with radial flow discussed in Sec.~\ref{s:pt_target}, localisation at the target requires $v_{\rm f} < \tilde{\kappa}/2$, whereas $\sigma_r^2$ diverges as $v_{\rm f}$ approaches this upper bound. 

Using Eq.~\eref{eq_density_uniform}, the entropy production rate associated with steering and the radial displacement variance are formally given by
\begin{equation}
\frac{\sigma_r^2}{\ell^2} = \frac{{\cal J}_3}{{\cal J}_1}~, \qquad
\frac{\ell^2 \dot{S}_c}{D_{\rm eff}} = \frac{{\cal J}_0}{{\cal J}_1}~,
\end{equation}
where ${\cal J}_n = \int_0^\infty\rmd x\, x^n e^{-x} I_0[2 x v_{\rm f} / \kapt + v_{\rm f} / \ell]$ and $I_0$ is the modified Bessel function of the first kind of order zero.
Although the symmetries of the problem do not allow to characterize the effect of the uniform flow via a single length scale, we show below that the escape of the particle at $v_{\rm f} \to \kapt/2$ is associated with similar characteristics of $\sigma_r^2$ and $\dot{S}_{\rm c}$ as those derived for the radial flow.

Using the expansion of the modified Bessel function $I_0(x) \simeq e^x/\sqrt{x}$ for $x \to \infty$, it is clear that for $v_{\rm f} \to \kapt/2$ the integrals ${\cal J}_n$ diverge as $ 1/(1 - 2v_{\rm f}/\kapt)^{n+\frac{1}{2}}$, such that
\begin{equation}
    \frac{\sigma_r^2}{\ell^2} \underset{v_{\rm f} \to \kapt/2}{\simeq} \left(1 - \frac{2 v_{\rm f}}{\kapt} \right)^{-2}~, \qquad 
    \frac{\ell^2 \dot{S}_c}{D_{\rm eff}} \underset{v_{\rm f} \to \kapt/2}{\simeq} 1 - \frac{2 v_{\rm f}}{\kapt}~.
\end{equation}
As for the radial flow, the escape of the particle coincides with a vanishing of the entropy production, together with a divergence of $\sigma_r^2$.
Moreover, the product $\sigma_r^2\dot{S}_c$ is found to diverge when $v_{\rm f} = \kapt/2$, highlighting again the tightening of the dissipation-accuracy tradeoff by the presence of a flow favouring particle escape.

\section{The discrete aligning policy: an exactly solvable model}\label{app:aligning_policy}
Here, we study a more minimal but fully solvable model of self-steering navigation by a smart active particle. Some of these results were included, albeit in a different context, in Ref.~\cite{cocconi2022statistical}. Similarly to Sec.~\ref{s:purely_align}, we assume that external flows are negligible, $\bm{v}(\bm{r})=0$, and consider an aligning policy which aims to localise the swimmer in the proximity of the $x$ coordinate axis, while inducing a net drift in the positive $x$ direction. The key simplification compared to the models studied in the main text, is to assume that the swimmer's heading angle is only allowed to take on the discrete values $\theta \in \{+\phi,-\phi\}$,  $\phi \leq \pi/2$, with Poisson switching rates controlled by the orthogonal distance from the target path. This essentially corresponds to the aligning policy (AP) introduced in Ref.~\cite{piro2022optimal}. The associated, adimensionalised Fokker-Planck equation reads
\begin{eqnarray}
\fl    \partial_t \bm{P}(\bm{r}) = - \partial_x[\cos(\phi)\bm{P} - {\rm Pe}^{-1} \partial_x \bm{P}] - \partial_y[\sin(\phi)\hat{T}\bm{P} - {\rm Pe}^{-1} \partial_y \bm{P}] + \hat{K}(y) \bm{P} ~,\label{eq:fp_discr}
\end{eqnarray}
with $\bm{P} \equiv (P_+,P_-)$ denoting the vector of conditional probability densities for a particular value $\theta = \pm \phi$ of the heading angle, $\hat{T} = {\rm diag}(1,-1)$ and
\begin{equation}
    \hat{K}(y) =  
    \tau \left[
\Theta(y)
\left(\begin{array}{cc}
-\epsilon & 1-\epsilon  \\
\epsilon & -1+\epsilon
\end{array}\right)
+ 
\Theta(-y)
\left(\begin{array}{cc}
-1+\epsilon & \epsilon  \\
1-\epsilon & -\epsilon
\end{array}\right)
\right]
\end{equation}
being the Markov matrix, where $\tau$ is the frequency of reorientation events and $0<\epsilon<1/2$ the error probability, while $\Theta(\cdot)$ denotes the Heaviside theta function. We further introduced the dimensionless Pecl{\'e}t number as ${\rm Pe}=w^2/(D_x \tau)$, rescaling $\tau=1$ without loss of generality. Motivated by the translational symmetry along the $x$ coordinate, we integrate Eq.~\eref{eq:fp_discr} with respect to the latter to obtain
\begin{equation}
    \partial_t \bm{P}(y) = - \partial_y[\sin(\phi)\bm{P} - {\rm Pe}^{-1} \partial_y \bm{P}] + \hat{K}(y) \bm{P} \label{eq:fp_discr_int}
\end{equation}
where $\bm{P}(y) \equiv \int dx \ \bm{P}(x,y)$ for compactness. Eq.~\eref{eq:fp_discr_int} can be solved analytically at steady-state by introducing the density $\Phi$ and polarity $\Psi$ fields
\begin{equation}
    \Phi(y) = \frac{1}{2}(P_-(y) + P_-+(y)), \quad \Psi(y) = \frac{1}{2}(P_-(y) - P_+(y))~.
\end{equation}
Indeed, by adding Eq.~\eref{eq:fp_discr} for $P_+$ and $P_-$ to each other, we obtain the current balance equation
\begin{equation}
    \sin(\phi) \Psi(y) = - {\rm Pe}^{-1} \partial_y \Phi(y) ~. \label{eq:current_balance}
\end{equation}
For ${\rm Pe} \to \infty$ this imposes $\Psi \to 0$, which in turn implies symmetric conditional distributions, $P_+(y) = P_-(y)$. Similarly, subtracting the two equations from each other and using \eref{eq:current_balance} we obtain a third order ODE for $\Phi$ only
\begin{equation} \label{eq:ode_lrnt_s}
    - \frac{1}{{\rm Pe}^2 \sin(\phi)} \partial_y^3 \Phi(y) + \left( \sin(\phi) + \frac{1}{{\rm Pe}\sin(\phi)} \right) \partial_y \Phi(y) + (1-2\epsilon) = 0~.
\end{equation}
Making an exponential ansatz for the solution we finally obtain 
\begin{eqnarray}
    \Phi(y) &= a_1 e^{-\alpha_1 |y|} + a_2 e^{-\alpha_2 |y|} \label{eq:phi_stst}\\
    \Psi(y) &= {\rm Pe}\sin(\phi)  {\rm sgn}(y) [a_1 \alpha_1 e^{-\alpha_1 |y|} + a_2 \alpha_2 e^{-\alpha_2 |y|} ] \label{eq:psi_stst}
\end{eqnarray}
with 
\begin{eqnarray}
    a_1 = \frac{-\alpha_1 \alpha_2^2}{4 (\alpha_1^2 -\alpha_2^2)}, \qquad a_2 = \frac{\alpha_1^2 \alpha_2}{4 (\alpha_1^2 -\alpha_2^2)}
\end{eqnarray}
the two integration constants, which we fix by enforcing normalisation, via $\int dy \ \Phi(y) = 1/2$, and continuity of the total probability density at $y=0$, via $\Psi(0)=0$. The coefficients $\alpha_{1,2}$ are the positive roots of the depressed cubic equation associated with \eref{eq:ode_lrnt_s}
\begin{equation} \label{eq:cubic_char_lrnt}
    \frac{1}{{\rm Pe}^{2}\sin(\phi)} \alpha^3 - \left( \sin(\phi) + \frac{1}{{\rm Pe}\sin(\phi)} \right) \alpha + (1-2\epsilon) = 0 ~.
\end{equation}
\begin{figure}
    \centering
    \includegraphics[width=\textwidth]{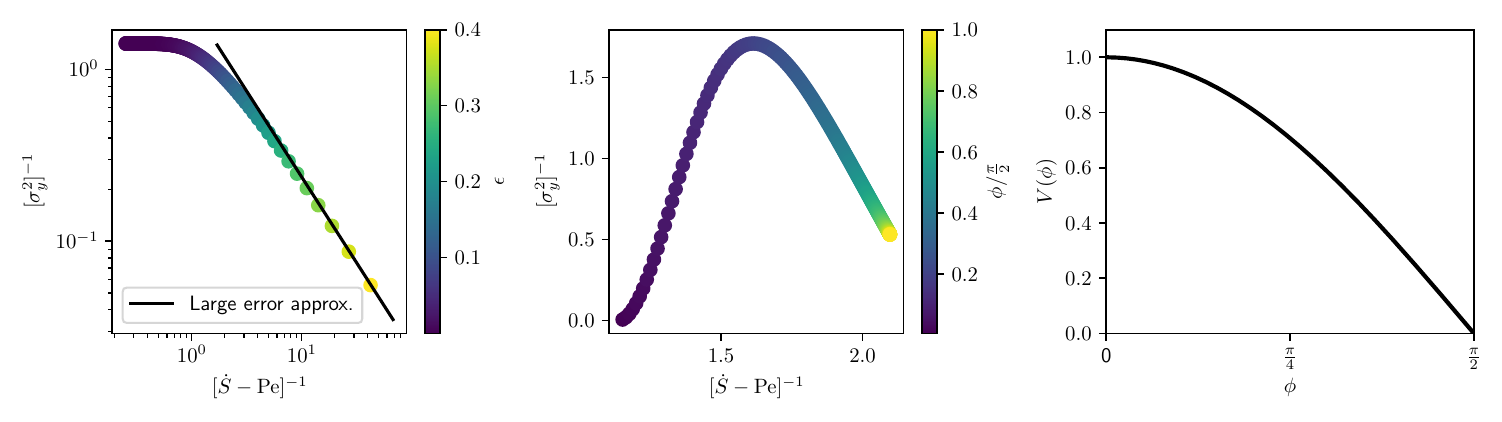}
    \caption{Adaptive contribution to the entropy production vs variance of the orthogonal displacements as a function of the error rate $\epsilon \in (0,1/2)$, left panel, and as a function of the heading angle $\phi \in (0,\pi/2]$, middle panel. The solid black line indicates the asymptotic scaling obtained from Eq.~\eref{eq:large_eps_scaling}. Note the divergence of the EPR at finite variance, associated with the approach to the limit $\epsilon \to 0$. Interestingly, the variance is a non-monotonic function of the heading angle. This is due to overshoot effects when adaptation occurs on a finite timescale. The right panel shows the longitudinal self-propulsion speed as a function of the orientation angle.}
    \label{fig:ap_curves}
\end{figure}
Compact expressions for these can be found in the limit of large error rate, where we write $\epsilon = (1-\delta \epsilon)/2$ with $\delta\epsilon \ll 1$. In particular,
\begin{eqnarray}
    \alpha_1 &= \frac{1}{\sin \phi  + (\sin \phi {\rm Pe})^{-1}} \delta\epsilon + \mathcal{O}(\delta\epsilon^2)~, \nonumber \\
    \alpha_2 &= {\rm Pe}\sqrt{\sin^2 \phi  + {\rm Pe}^{-1}} - \frac{1 }{2[\sin \phi  + (\sin \phi {\rm Pe})^{-1}]} \delta\epsilon + \mathcal{O}(\delta\epsilon^2)~.\label{eq:roots_smalleps}
\end{eqnarray}
Equipped with Eqs.~\eref{eq:phi_stst} and \eref{eq:psi_stst}, we can now compute observables of physical interest in closed form, such as the variance of the orthogonal displacement
\begin{equation}
    \sigma_y^2 = \int dy \ y^2 (P_+(y) + P_-(y)) = \frac{1}{\alpha_1^2} + \frac{1}{\alpha_2^2}
\end{equation}
and the rate of entropy production \cite{cocconi2020entropy}
\begin{eqnarray}
    \dot{S} 
    &= {\rm Pe} + \int dy \ (P_+ K_{+-} - P_- K_{-+}) \log \frac{P_+ K_{+-}}{P_- K_{-+}} \nonumber \\
    &= {\rm Pe} + \frac{1}{4} \left( \log \frac{1-\epsilon}{\epsilon} \right) \left( (1-2\epsilon) + \frac{1}{{\rm Pe} \sin(\phi)} \frac{\alpha_1 \alpha_2}{\alpha_1 + \alpha_2} \right) ~,\label{eq:epr_arnt}
\end{eqnarray}
where the matrix element $K_{ab}$ denotes the rate of Poisson switching from self-propulsion velocity $a$ to $b$.
For the mean longitudinal speed, we have here trivially $V = \langle \cos\theta \rangle = \cos\phi$. We can see from \eref{eq:epr_arnt} that the dissipation diverges logarithmically in the limit of irreversible switching, $\epsilon \to 0$. In the large error limit, Eq.~\eref{eq:roots_smalleps}, the variance and entropy production are given approximately as 
\begin{equation}\label{eq:large_eps_scaling}
    \sigma_y^2 = \left[ \sin\phi + (\sin\phi{\rm Pe})^{-1} \right]^2 \delta\epsilon^{-2}, \quad \dot{S}_{\rm c} = \frac{1}{2}\left( 1 + \frac{1}{1 + \sin^2\phi{\rm Pe}} \right)\delta\epsilon^{2}~,
\end{equation}
whereby the product $\sigma_y^2 \dot{S}_{\rm c}$ is independent of $\delta \epsilon$. The same power law relationship was observed in Sections.~\ref{s:pt_target} and~\ref{ss:aap} (cf.~Figs.~\ref{fig:point_target} and~\ref{fig:AAP}) and highlights once again the existence of a fundamental dissipation-accuracy tradeoff. 



\section{Purely compensating policies}\label{a:cp}

To isolate the contribution to the thermodynamic dissipation associated solely with adaptation to, and compensation for, an external flow, we consider here policies $\hat{\bm{u}}^*(\bm{v}(\bm{r}))= (\cos\theta^*,\sin\theta^*)$ which depend on the local orthogonal flow velocity component but not on the orthogonal displacement from the target subspace. See again Fig.~\ref{fig:quiver_policies}(a) for schematic illustration.  The simplest such \emph{purely compensating} policy (CP) is given by
\begin{equation}\label{eq:cp_simplest}
    \sin\theta^*(\bm{r}) = -v_y(\bm{r})~, \quad \cos\theta^*(\bm{r}) = \sqrt{1-\sin^2\theta^*}~.
\end{equation}
Under ideal conditions, where adaptation is perfect and instantaneous, this steering policy should result in an exact cancellation of the advective contribution to the orthogonal displacement, such that the net dynamics of the latter become purely diffusive with bare diffusivity coefficient ${\rm Pe}^{-1}$, while those of the longitudinal displacement are drift-diffusive. 

\begin{figure}
    \centering
    \includegraphics[width=\textwidth]{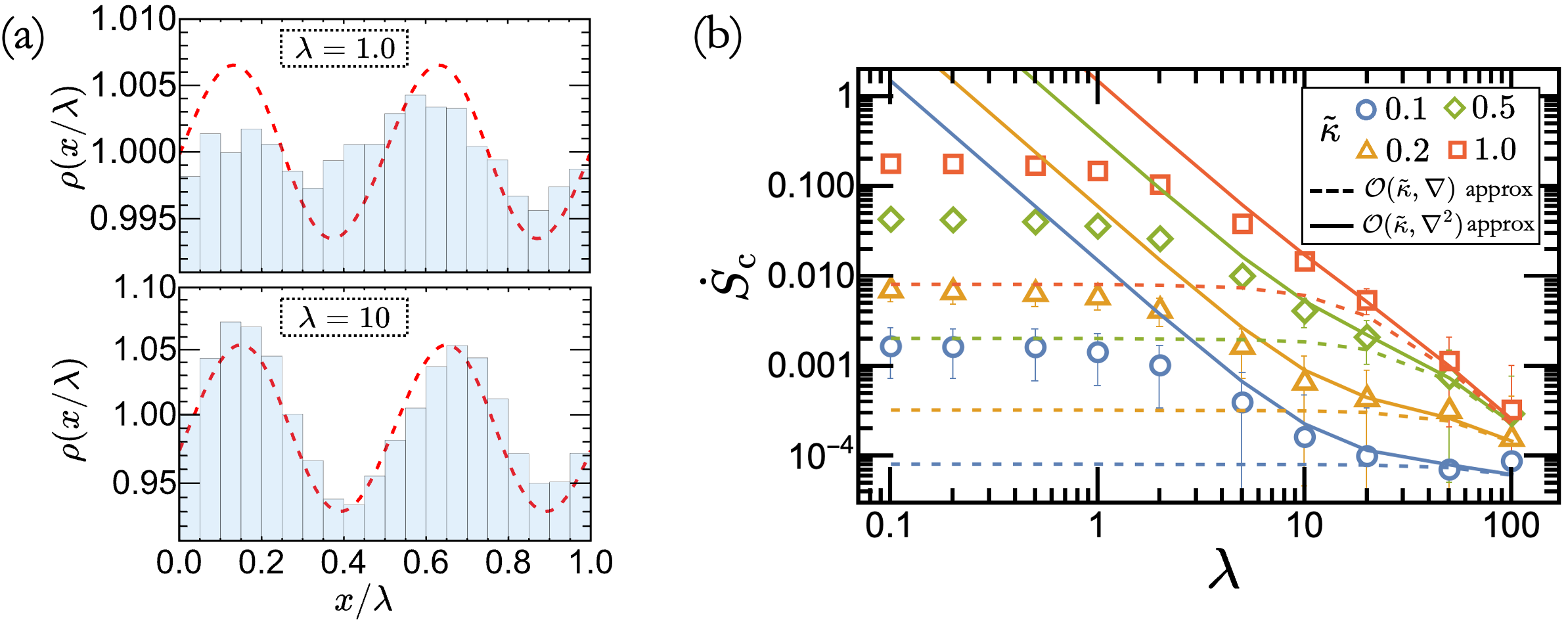}
    \caption{Comparison between theory and simulations of the purely compensating policy (CP) defined in Eq.~\eref{eq:cp_simplest} with a sinusoidal flow of wavelength $\lambda$. (a) Empirical histograms of the longitudinal displacement obtained from numerical simulations are compared to the theoretical prediction~\eref{eq:sol_dens_driv}, dashed red curves, for two different values of the flow wavelength $\lambda$. Consistently with the range of validity of the gradient expansion, we observe good agreement only for large $\lambda$. (b) Numerical results versus theoretical predictions, showing both approximations $\mathcal{O}(\kapt,\nabla)$, dashed curves, and $\mathcal{O}(\kapt,\nabla^2)$, solid curves, for the entropy production for four different values of $\kapt \lesssim 1$ as indicated by the legend. All data shown in (a) and (b) have been obtained at $\Pe = 10$ and for a flow amplitude $A = 0.8$ in units of the self-propulsion speed.}
    \label{fig:CP_plots}
\end{figure}

For the sake of analytical tractability, we focus our analysis on the same orthogonal sinusoidal flow introduced in Sec.~\ref{s:caap}, such that $\bm{v}(\bm{r}) = (0,v_y(x))$ with $v_y(x) = v_y(x+\lambda) = \tilde{v}_y[(x/\lambda) \ {\rm mod} \ 1 ]$ and $\lambda$ the characteristic length scale. Thus, $\sin\theta^*(\bm{r})$ in~\eref{eq:cp_simplest} is also periodic with period $\lambda$.
In this framework, the problem becomes translationally invariant with respect to the $y$ coordinate. Within the gradient expansion, we can thus integrate the density dynamics \eref{eq:cg_fp_rho_main} to obtain an approximate Fokker-Planck equation for the longitudinal displacement
\begin{equation}\label{eq:fp_long_comp}
    \partial_t \varrho(x) = D_{\rm eff} \partial_x^2 \varrho - \frac{\kapt}{2} \partial_x [\varrho \cos\theta^*(x) ]~,
\end{equation}
where we introduced the shorthand notation $\varrho(x) \equiv \sum_{n \in \mathbb{Z}}\int \rmd y \varrho(x + n\lambda,y)$, $x \in [0,\lambda)$. 
Equation~\eref{eq:fp_long_comp} can be interpreted as the Fokker-Planck equation for a driven-diffusion process on a ring of length $\lambda$ with diffusivity $D_{\rm eff}$ and constant drift $\mu \equiv \kapt\int_0^\lambda \rmd x  \cos\theta^*(x)/2$, in the presence of a periodic potential $\kapt W(x)/2$ satisfying $W'(x) \equiv 2\mu/\kapt - \cos\theta^*(x) $ \cite{risken1996fokker}. 
Further demanding that $\cos\theta^*(x)$ is a function of $x/\lambda$ only, implies 
\begin{equation}\label{eq:pot_dep_lambda}
    W(x;\lambda)\equiv\lambda \tilde{W}(x/\lambda)~.
\end{equation}
The steady-state solution of this driven problem is \cite{risken1996fokker}
\begin{equation}\label{eq:sol_dens_driv}
\fl \qquad\qquad   \varrho(x;\lambda) = \mathcal{N} {\rm exp}\left[ \frac{\int^x \rmd x' \cos\theta^*(x')}{\ell} \right] \int_x^{\lambda+x} \rmd s \ {\rm exp}\left[ -\frac{\int^s \rmd x' \cos\theta^*(x')}{\ell}\right]~,
\end{equation}
with $\mathcal{N}^{-1} = \int \rmd x \varrho(x)$ a normalisation factor and $\ell = 2D_{\rm eff}/\kapt$. For large $\lambda$, this result is found to be in good agreement with empirical histograms obtained from numerical simulations, see Fig.~\ref{fig:CP_plots}(a).
Using Eq.~\eref{eq:sdot_c_11mom} for the rate of entropy production to $\mathcal{O}(\kapt,\nabla)$ together with Eq.\eref{eq:sol_dens_driv}, the dissipation $\dot{S}_{\rm c}$ can be written as
\begin{eqnarray}\label{eq:scaling_EPR_CP}
    \dot{S}_{\rm c} &=    \frac{\kapt}{2} \int_0^1 \rmd s \ (\partial_s \cos\theta^*(\lambda s;\lambda))\varrho(\lambda s;\lambda) + \mathcal{O}(\kapt^2,\nabla^2)~.
\end{eqnarray}
where we have rescaled the integral using $s = x/\lambda$.
Importantly, as long as $\theta^*$ depends on $x$ only via $x/\lambda$, such that $\sin\theta^*(\lambda s;\lambda) = \sin \theta^*(s;1)$ for $s\in[0,1)$, the integral above depends on $\lambda$ only via $\varrho(\lambda s;\lambda)$. It follows that the asymptotic dependence of $\dot{S}_{\rm c}$ on $\lambda$ can be obtained by expanding $\varrho(\lambda s;\lambda)$ as a power series and simply identifying the order of the leading, non-vanishing contribution.

Let us first consider the regime $\lambda \to \infty$ of Eqs.~\eref{eq:sol_dens_driv} and \eref{eq:scaling_EPR_CP}. We can argue based on~\eref{eq:pot_dep_lambda} that in this limit diffusion is negligible.
Also, since $\cos\theta^* > 0$ for all $x$ by construction of the policy, the driven-diffusion problem is always in the ``running'' (as opposed to ``locked'') phase \cite{risken1996fokker}. Consequently, we expect that the probability density will be approximately proportional to the reciprocal of the local drift velocity
\begin{equation}\label{eq:rho_large_lambda}
    \varrho(\lambda s;\lambda) = \frac{\lambda^{-1} \cos^{-1}[\theta^*(s;1)]}{\int_0^1 \rmd s \cos^{-1}\theta^*(s;1) } + \mathcal{O}(\lambda^{-2})~.
\end{equation}
Substitution of Eq.~\eref{eq:rho_large_lambda} into~\eref{eq:scaling_EPR_CP} yields a vanishing $\dot{S}_c$ to leading order in large $\lambda$. The first non-zero contribution must thus decay at least as fast as $\lambda^{-2}$. This result is in agreement with what we might have expected on physical grounds, since a slowly varying policy requires less frequent adaptation.

As for the opposite regime, $\lambda \to 0$, we cannot rely on Eq.~\eref{eq:pot_dep_lambda} to determine the marginal density $\varrho(x)$ since the gradient expansion is expected to break down for large flow gradients.
%
Sill, we can argue that, in this regime, $\theta^*$ should decorrelate infinitely fast as a consequence of thermal diffusion, leading to effective $\lambda$-independent angular dynamics $\dot{\theta}(t) = - \tilde{\kappa}\langle \cos\theta^*\rangle_{\theta^*} \sin\theta + \sqrt{2}\xi_\theta(t)$, with $\langle \cdot \rangle_{\theta^*}$ indicating a spatial average with respect to $P(\theta^*) = \lambda^{-1}/\partial_x\theta^*(x)$ on $x \in [0,\lambda]$. 
We then have, via Eq.~\eref{eq:scaling_EPR_CP}, that $\dot{S}_{\rm c} \sim \lambda^0$ to leading order in small $\lambda$, in qualitative agreement with observations made in Fig.~\ref{fig:quiver_policies}(a). 
Analogous results, where dissipation becomes independent of the potential correlation timescale as the latter approaches zero, have previously been obtained for Brownian motion in a fluctuating potential \cite{alston2022non}. This has sometimes been referred to as entropic anomaly \cite{bo2014entropy,celani2012anomalous}.

Lastly, we shall now compare our theoretical prediction concerning the rate of entropy production of CP, obtained by combining Eqs.~\eref{eq:sol_dens_driv} and \eref{eq:scaling_EPR_CP}, with direct measurements from our Langevin simulations. As shown in Fig.~\ref{fig:CP_plots}(b), we observe quantitative agreement between the two at large $\lambda$ (note that the comparison is done without fitting parameter), while $\dot{S}_{\rm c}$ is typically underestimated by the theory in the small $\lambda$ regime.
Pushing the gradient expansion to the next order (see Eq.~\eref{eq:epr_corr_k1_n2_app_CP} and details in~\ref{app_Onabla2_exp}), we nonetheless find that the range of quantitative validity of the theory can be extended by around one order of magnitude in $\lambda$.

\vspace{1cm}

\bibliographystyle{iopart-num}
\bibliography{updated_bibliography}

\end{document}